\documentclass{sig-alternate}

\usepackage{enumitem}
\usepackage{framed}
\usepackage{cprotect}
\usepackage{enumitem}
\usepackage{listings}
\usepackage{amstext}
\usepackage{amstext}
\usepackage{pdfpages}
\usepackage{alltt}
\usepackage{epstopdf}
\usepackage{xspace,colortbl}
\usepackage[USenglish]{babel}
\usepackage{multirow}
\usepackage[hyphens]{url}
\usepackage{subfigure}
\usepackage{graphicx}
\usepackage{amssymb}
\usepackage{fmtcount}
\usepackage{amsfonts}
\usepackage{xspace}
\usepackage{amsmath}
\usepackage{multirow}
\usepackage[mathscr]{eucal}
\usepackage{colortbl}

\usepackage{amsmath,amssymb}
\usepackage[linesnumbered, ruled,vlined]{algorithm2e}

\usepackage{caption}
\usepackage{graphicx}

\usepackage{bm}
\usepackage{times}
\usepackage[nospace]{cite}
\usepackage{csquotes}
\usepackage{enumitem}

\usepackage{balance}

\makeatletter
\def\@copyrightspace{\relax}
\makeatother

\DeclareMathOperator*{\argmax}{arg\,max}

\begin{document}


\newtheorem{theorem}{Theorem}
\newtheorem{example}{Example}
\newtheorem{definition}{Definition}
\newtheorem{problem}{Problem}
\newtheorem{property}{Property}
\newtheorem{proposition}{Proposition}
\newtheorem{lemma}{Lemma}
\newtheorem{corollary}{Corollary}

\newcommand{\detectlib}{\texttt{IsoDetect}\xspace}
\newcommand{\company}{\texttt{Company X}\xspace}
\newcommand{\cond}{\textrm{pred}\xspace}
\newcommand{\dataset}{data set\xspace}
\newcommand{\datasets}{data sets\xspace}
\newcommand{\spview}{\textsf{SPView}\xspace}
\newcommand{\fjview}{\textsf{FJView}\xspace}
\newcommand{\aggview}{\textsf{AggView}\xspace}
\newcommand{\hashfunc}[1]{\textsf{hash}(#1)\xspace}
\newcommand{\hashop}{\textsf{hash}\xspace}
\newcommand{\nsc}{\textsf{NormalizedSC}\xspace}
\newcommand{\rsc}{\textsf{RawSC}\xspace}

\newcommand{\avgfunc}{\ensuremath{\texttt{avg} }\xspace}
\newcommand{\maxfunc}{\ensuremath{\texttt{max} }\xspace}
\newcommand{\minfunc}{\ensuremath{\texttt{min} }\xspace}
\newcommand{\histfunc}{\ensuremath{\texttt{histogram\_numeric} }\xspace}
\newcommand{\countfunc}{\ensuremath{\texttt{count}}\xspace}
\newcommand{\sumfunc}{\ensuremath{\texttt{sum} }\xspace}
\newcommand{\varfunc}{\ensuremath{\texttt{var} }\xspace}
\newcommand{\stdfunc}{\ensuremath{\texttt{std} }\xspace}
\newcommand{\covfunc}{\ensuremath{\texttt{cov} }\xspace}
\newcommand{\corrfunc}{\ensuremath{\texttt{corr} }\xspace}
\newcommand{\medfunc}{\ensuremath{\texttt{median} }\xspace}
\newcommand{\percfunc}{\ensuremath{\texttt{percentile} }\xspace}
\newcommand{\havingfunc}{\ensuremath{\texttt{HAVING} }\xspace}
\newcommand{\selectfunc}{\ensuremath{\texttt{select} }\xspace}
\newcommand{\ratio}{\ensuremath{\rho }\xspace}

\newcommand{\insertion}{\ensuremath{\texttt{INSERT} }\xspace}
\newcommand{\update}{\ensuremath{\texttt{UPDATE} }\xspace}
\newcommand{\delete}{\ensuremath{\texttt{DELETE} }\xspace}

\newcommand{\sysfull}{BoostClean\xspace}
\newcommand{\sys}{BoostClean\xspace}
\newcommand{\sysnospace}{BoostClean}

\newcommand{\tbl}[1]{\textsf{#1}\xspace}
\newcommand{\field}[1]{\textsf{#1}\xspace}
\newcommand{\cost}{\textrm{cost}\xspace}
\newcommand{\ans}{\textsf{ans}\xspace}
\newcommand{\dans}{\Delta\textsf{ans}\xspace}
\newcommand{\cqp}{correction query processing\xspace}
\newcommand{\Cqp}{Correction query processing\xspace}

\newcommand{\reminder}[1]{{{\textcolor{magenta}{\{\{\bf #1\}\}}}\xspace}}
\newcommand{\ewu}[1]{{{\textcolor{blue}{\{\{\bf ewu:\} #1\}}}\xspace}}
\newcommand{\mps}[1]{{{\textcolor{red}{\{\{\bf meelap:\} #1\}}}\xspace}}
\newcommand{\stitle}[1]{\vspace{0.5em}\noindent\textbf{#1}}

\definecolor{light-gray}{gray}{0.95}
\definecolor{mid-gray}{gray}{0.85}
\definecolor{green}{RGB}{0,176,80}
\definecolor{darkred}{rgb}{0.7,0.25,0.25}
\definecolor{darkgreen}{rgb}{0.15,0.55,0.15}
\definecolor{darkblue}{rgb}{0.1,0.1,0.5}
\definecolor{orange}{RGB}{237,125,49}
\definecolor{blue}{RGB}{68,114,196}
\definecolor{pop}{RGB}{0,21,245}

\newcommand{\blue}[1]{{\textcolor{blue}{{\bf #1}}\xspace}}
\newcommand{\orange}[1]{{\textcolor{orange}{{\bf #1}}\xspace}}
\newcommand{\pop}[1]{{\textcolor{pop}{{\textit{\textbf{#1}}}}\xspace}}

\newcommand{\specialcell}[2][c]{%
  \begin{tabular}[#1]{@{}c@{}}#2\end{tabular}}

\def\ojoin{\setbox0=\hbox{$\bowtie$}%
  \rule[-.02ex]{.25em}{.4pt}\llap{\rule[\ht0]{.25em}{.4pt}}}
\def\leftouterjoin{\mathbin{\ojoin\mkern-5.8mu\bowtie}}
\def\rightouterjoin{\mathbin{\bowtie\mkern-5.8mu\ojoin}}
\def\fullouterjoin{\mathbin{\ojoin\mkern-5.8mu\bowtie\mkern-5.8mu\ojoin}}

\pagestyle{plain}

\title{\sys: Automated Error Detection and Repair for Machine Learning}

\numberofauthors{1}
\author{ Sanjay Krishnan$\,^{*}$, Michael J. Franklin$\,^{*\dag}$, Ken Goldberg$\,^{*}$, Eugene Wu{$\,^{\dag\dag}$}  \\
\affaddr{ $^*$UC Berkeley, ~~ $^\dag$University of Chicago, ~~ $^{\dag\dag}$Columbia University} \\
\affaddr{ \{sanjaykrishnan, franklin, goldberg\}@berkeley.edu ~~ ewu@cs.columbia.edu}\\
\affaddr{}
}

\maketitle

\begin{abstract}
Predictive models based on machine learning can be highly sensitive to data error.
Training data are often combined from a variety of different sources, each susceptible to different types of inconsistencies, and as new data stream in during prediction time, the model may encounter previously unseen inconsistencies.
An important class of such inconsistencies are {\it domain value violations} that occur when an attribute value is outside of an allowed domain.
We explore automatically detecting and repairing such violations by leveraging the often available clean test labels to determine whether a given detection and repair combination will improve model accuracy.
We present \sys which automatically selects an ensemble of error detection and repair combinations using statistical boosting.
\sys selects this ensemble from an extensible library that is pre-populated general detection functions, including a novel detector based on the Word2Vec deep learning model, which detects errors across a diverse set of domains.
Our evaluation on a collection of 12 datasets from Kaggle, the UCI repository, real-world data analyses, and production datasets that show that \sys can increase absolute prediction accuracy by up to 9\% over the best non-ensembled alternatives.
Our optimizations including parallelism, materialization, and indexing techniques show a $22.2\times$ end-to-end speedup on a 16-core machine.  
\end{abstract}


\section{Introduction}\label{intro}\sloppy
The availability of data and vast cloud-based computational resources has ushered in an era of more sophisticated machine learning (ML) models in prediction, recommendation, and automation.
The database community has built systems to support almost every stage of the development process including featurization~\cite{keystone,zhang2014mat}, distributed model training~\cite{hellerstein2012madlib, crotty2014tupleware, feng2012towards, tensor}, and model deployment~\cite{crankshawmissing}.  
However, an under-served, yet crucial, component is the management and cleaning of dirty data. 
 If unaccounted for, this dirty data can drastically bias predictions that are undesirable or even dangerous~\cite{vanderbilt2012let}.  Recent papers and surveys of analysts suggest that such problems are pervasive~\cite{sculley2014machine,kandel2012,krishnan2016hilda}.

As a concrete example, we are collaborating with a data science company called \company\footnote{Anonymized at the request of the company.} that ranks sales leads based on Salesforce.com data on past sales leads, and additional information scraped from the web about the client.
The company predicts the probability of viability for a potential (unlabeled) lead.
The data are acquired from a combination of manual data entry and automatically scraped web sources, and thus, inconsistencies, missing data, and incorrect values are a significant problem.  For instance, a typical error is the inconsistent representation of missing values (e.g., ``-999'', ``EMPTY'' or ``none'' may be used depending on the sales representative).  If the featurization code does not recognize and address these errors, it can lead to biases that degrade the quality of the model. For example, the data scientist may impute a default mean value for all blank attributes but miss the code ``-999'', which is then interpreted as a semantic value. 
Detecting and repairing all such errors is extremely time-consuming, and for every new client this effort will have to be repeated.

This company's data cleaning challenges are not unique and are prevalent in many industrial ML pipelines~\cite{krishnan2016hilda}.  
Software Engineers write custom conditional cleaning scripts that are a combination of a {\it detector}, which are a collection of Boolean functions that specify a subset of records that are dirty, and {\it repair} functions that transform or delete those records.  It is not enough to write these scripts once.
The predictive nature of ML applications means that the system will continuously encounter and process new, unseen data.
Software Engineers must constantly monitor and maintain the data processing pipeline to account for unexpected changes~\cite{sculley2014machine, DBLP:conf/sigmod/KrishnanFGWW16}.
To further exacerbate this problem, modern prediction models rely on data integrated from a wide variety of sources (e.g., \company combines on average 5-10 sources to train a model).  For each data source, the engineer must understand domain-specific information (e.g., invariants) in order to accurately clean the data.  For instance, we found that each machine learning dataset required between $1-7$ custom error detection rules in order to identify the low-hanging errors in those datasets.

To reduce this burden, we present a new system, called \sys, that automates the process of detecting and repairing a common class of data errors called {\it domain value violations} that occur when an attribute value is outside of its value domain.  Numerous data quality surveys across the database, statistics, and scientific literature highlight the prevalence, variety, and importance of this class of errors, which include missing data, incorrect data, or inconsistent representations of the same logical data value~\cite{muller2005problems,li2010improving,kim2003taxonomy,kandel2011research}.  \sys focuses on this common class of errors, and leaves more complex scenarios such as entity resolution to the Software Engineers. After deployment, \sys can help ensure that deployed models maintain high accuracy even in the presence of incoming dirty data, and engineers are only needed to address drastic changes to the input data. 

In traditional relational data cleaning, it is very hard to quantify the accuracy of an automatic data cleaning process without ground truth--a dataset where {\it all attributes are fully correct}.
On the other hand, in ML, cleanly labeled test data is often available (e.g., the results of following a sales lead). 
Labels often represent directly observed phenomena making them relatively clean, while features are often weaker signals integrated from multiple disparate sources and subject to error and frequent change.
This allows us to define accuracy in terms of the model's predictive accuracy--the data cleaning being a means to improving that predictive accuracy.
In this sense, our goal is not to fully clean each record and recover a consistent relation; instead, to utilize the available cleaning resources to best improve a model trained on this dataset.
The key challenge is to efficiently search the space of possible conditional data cleaning scripts (detector and repair combinations) while ensuring that the model does not overfit~\cite{DBLP:journals/pvldb/KrishnanWWFG16,krishnan2016hilda}.   

Our primary observation is that a conditional cleaning script can be interpreted as generating a new set of features (the cleaned values), and thereby generating a new model trained on those cleaned features. 
We can view the process of selecting the best sequence of cleaning operations as an ensembling problem, i.e., selecting the best collection models that collectively estimate a label. 
Although there are many possible algorithms~\cite{dietterich2000ensemble}, we use a powerful technique called Boosting~\cite{freund1995desicion} that composes a set of weak learners into a strong learner.  
First, unlike methods that are specific to certain classes of models (e.g., linear models, differentiable models), boosting can be applied to black-box models. 
Second, it takes interactions and correlations between the different data cleaning models into account by incrementally selecting ``orthogonal'' compositions.

\sys takes as input a relational table, a library of detector functions $\mathcal{D}$ that generate (possibly incorrect) predicates that match candidate dirty records, a library of repair functions $\mathcal{F}$ that transform or delete a record, and a user-specified classifier training procedure \texttt{train()}.
\sys has two key components: an automatic error detector to determine subsets of records that are dirty, and a repair selector to select repair actions for those dirty records using boosting.
We cast the former component into a featurization problem so that the user focuses on the familiar task of creating feature extraction functions while \sys translates these features into error detection rules using a technique called Isolation Forests~\cite{liu2008isolation}.  Further, we have written a general set of featurizers, including one that is a novel adaptation of the \textsf{word2vec} neural network architecture that is effective at detecting multi-attribute errors.  The neural network can be individually tailored to each dataset and learn to predict the co-occurrence of attributes in a record. 
The detectors output relational predicates $p_i$, which can be used to detect candidate errors.  The second component then uses boosting to generate a sequence of conditional cleaning scripts $(p_i, r_i)$ to be applied to the training and test datasets, where $r_i$ is the repair function to be applied to records matching predicate $p_i$.

This paper focuses on data errors that cause domain value violations in the context of supervised classification models (both single and multi-class).  The system is currently designed for a single-node setting. Our contributions are as follows:

\vspace{0.25em}\noindent\textbf{Cleaning as Boosting: } We present a new automated data cleaning system based on statistical boosting that finds the best ensemble of operations from a library of operations to maximize the predictive performance of a downstream model. 

\stitle{Automatic Model Improvements:} We evaluated \sys on 12 datasets collected from Kaggle, the UCI repository, real-world data analyses, and \company, and improved absolute prediction accuracy by up to $9\%$ in comparison to baseline (non-ensembled integrity constraint+quantitative outlier detector) approaches on completely unseen test data. 

\vspace{0.5em}\noindent\textbf{Error Detection Library: } We have built an optimized library of data cleaning operations based on deterministic rules and statistical criteria from which \sys selects. To better detect errors in categorical attributes, we developed a novel detector based on the \textsf{Word2Vec} neural network architecture. Following prior experimental procedures~\cite{DBLP:journals/pvldb/AbedjanCDFIOPST16}, the library achieves a detection accuracy of 81\% of all of the errors found by hand-written rules on eight machine learning datasets.  

\vspace{0.5em}\noindent\textbf{Optimizations: } Our optimizations including parallelism, materialization, and indexing techniques show a $22.2\times$ end-to-end speedup on a 16-core machine.

\section{Background}
This section motivates \sys in relation to prior work.
Using the pilot study as inspiration, we use a simplified running example to present the system and notation:
\begin{example}[Lead Prediction]\sloppy\label{ex:lead}
Past clients are stored in a relational database:
\[
R(id, name, num\_emp, industry, region, successful)
\]
where $name$ is the company name, $num\_emp$ is the number of employees in the company, $industry$ is a categorical attribute that describes the industry segment, $region$ is a code indicating the region of the country the business is headquartered, and $is\_successful$ is a Boolean describing whether the company purchased the product.
\end{example}

\begin{figure}[t]
\centering
 \includegraphics[width=\columnwidth]{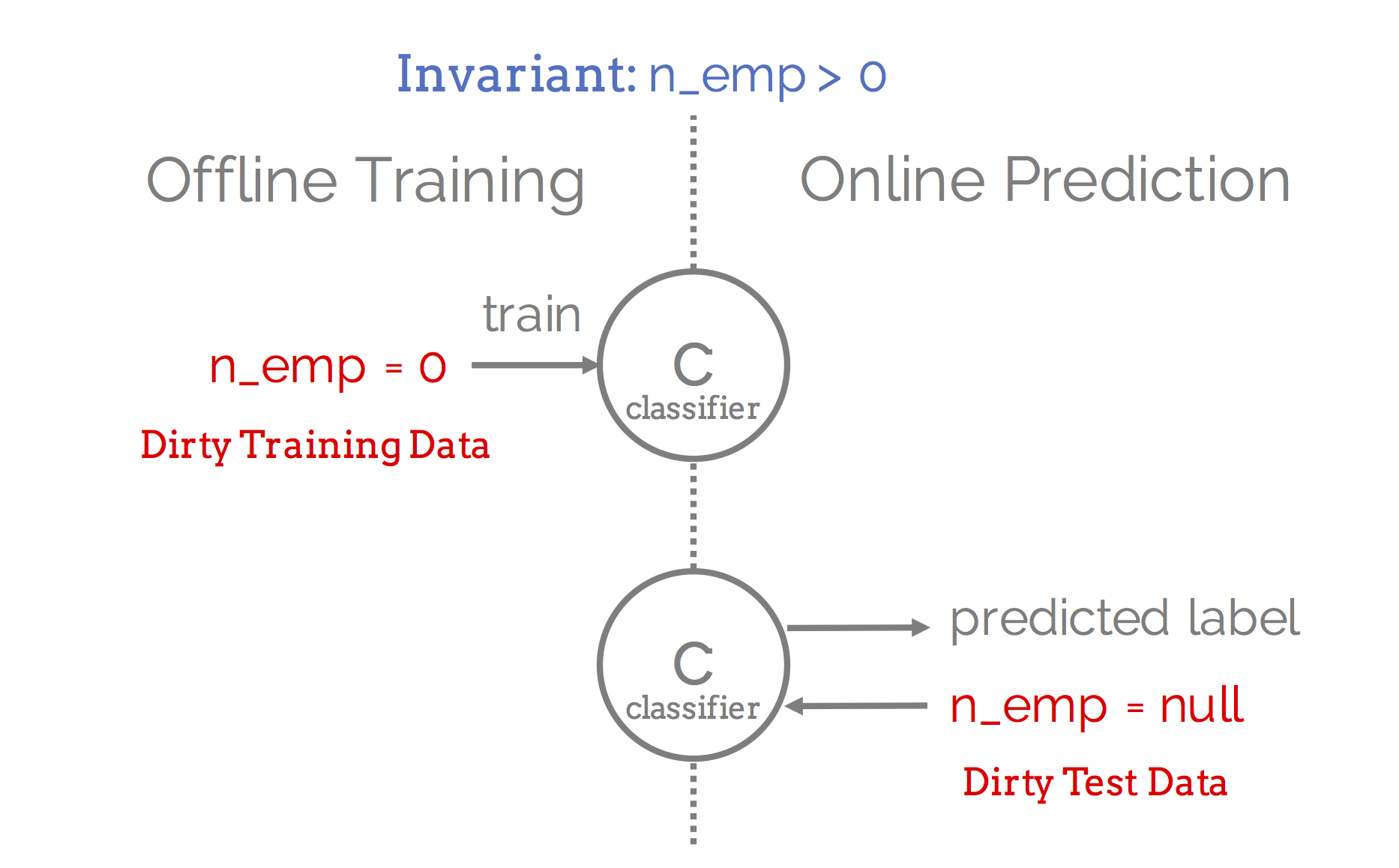}
 \caption{The above example uses the invariant that the number of employees is greater than $0$.  Training errors are violations of the invariant in the training dataset (top row). Prediction errors are invariant violations in the test data during prediction. \sys is a tool to {\it detect} both types of errors and generate a corrective {\it repair} action.
 \label{fig:error}}
\end{figure}

\subsection{Machine Learning and Dirty Data}
It is important to highlight a number of points when considering data cleaning in the machine learning context.
First, machine learning models are often robust to statistical noise and inherent variations in the dataset.  For this reason, our focus is not to reduce random noise; instead, our focus is to identify and address systematic errors due to invariant violations that lead to unforeseen biases in the model.  
For example, records with a positive classification label are more likely to have NULL values in the training set.
Figure \ref{fig:error} illustrates two examples of how this can affect a model.
Second, while the notion of invariants is similar to integrity constraints in traditional relational data cleaning, the way that repairs are evaluated differ.  Relational data cleaning focuses on identifying a minimal set of repairs that resolve a set of constraint violations.  On the other hand, our goal is to improve the the downstream model test prediction accuracy.  These goals are not necessarily aligned, and by ignoring the downstream model, it is possible that traditional cleaning techniques perform cleaning operations that degrade the model accuracy.

\subsection{Existing Approaches}
\sys brings together generic, dataset-independent error detection with automatically learned repair strategies for ML applications.
We review how baseline techniques proposed in prior work that could apply to this problem.

\vspace{0.5em}

\noindent\textbf{Rule-based Repair: } 
In the running example, suppose some of the values for $num\_emp$ are NULL and we want to train a classifier to predict $is\_successful$.
We would define a domain integrity constraint $num\_emp \ne~NULL$, and then propose a set of repairs to satisfy this constraint.
With no other information, this rule-based approach could in principle impute any non-null value from the domain to create a logically consistent relation.
To avoid this problem, we can adopt an approach like~\cite{prokoshyna2015combining} and select the imputations that minimize the statistical distance of the updated relation to an ideal distribution for the attribute, for example, an ideal power-law distribution.
This would impute values in such a way that $num\_emp$ matched a Zipfian distribution.

When we train a classifier after applying such a technique, counter-intuitive effects can occur.  
The data cleaning operation may break important correlations in the data and may introduce biases into the training data not present in test conditions. 
Consider the degenerate case where $num\_emp = NULL$ is perfectly correlated with one of the prediction classes--in this case, it may be better to NOT clean the data!
While more sophisticated statistical imputation techniques exist~\cite{schafer1998multiple}, they all have the same fundamental problem that the value imputation is divorced from the downstream classifier's predictive accuracy. 
We see this problem in our experiments (Section~\ref{exp:comp}), where on some datasets imputing the most frequent value leads to a more accurate downstream classifier than imputing to minimize the difference from an ideal distribution.

\vspace{0.5em}\noindent\textbf{Statistical Detection: } Rule-based techniques are dependent on the analyst defining the invariants. Defining such invariants can be challenging if the analyst is working with a new dataset, if she cannot anticipate how future data might look, or if the number of datasets is too large (e.g., in a data lake setting).
There is a well-established line of literature on statistical anomaly detection~\cite{hellerstein2008quantitative}, and for the most part, these techniques are generic and dataset independent (up-to hyperparameters). Typically, such approaches identify \emph{outlier} records outside of some normal range of variance. However, the problem is that not all dirty data look like outliers. In the running example, there could truly be companies where $num\_emp = 0$. It has been shown that statistical anomaly detection techniques miss obvious errors in heterogeneous datasets that contain a mixture numerical, categorical, and string-valued attributes~\cite{DBLP:journals/pvldb/AbedjanCDFIOPST16}.

Abedjan et al. recently evaluated a wide range of error detection techniques on 5 proprietary real-world datasets~\cite{DBLP:journals/pvldb/AbedjanCDFIOPST16}.  They found that the errors in 3 of the datasets were dominated by missing cell values; 1 dataset contained functional dependency violations due to erroneous numerical attribute values, and only 1 dataset require complex user-specified denial constraints~\cite{chu2013discovering} to identify the errors.  These findings suggest that, in 4 of the 5 datasets, a significant portion of data errors can be classified as {\it domain integrity errors}, wherein a cell contains a value outside of its domain of permissible values.

\stitle{Towards Automated Cleaning:} We believe this highlights the potential value of automated cleaning systems such as \sys to identify the bulk of common-case errors, so that developers may focus on the specialized, domain-specific errors.  The prevalence of {\it domain integrity errors} suggests that a pre-defined set of featurizers and detector generators can be sufficient to detect these errors.  In fact, on 8 of our experimental datasets, \sys using our pre-populated detector library achieves a detection accuracy of 81\% of all of the errors found by hand-written rules.

Therefore, we need a mix of statistical rules and logic rules to determine errors.
We explore to what extent we can derive these rules from data for routine errors. 
We surveyed 8 ML datasets used in Kaggle competitions and benchmarks in the UCI ML repository, and found that a majority of the non-statistical errors could be detected as \emph{domain integrity constraints}, i.e., disallowed values in single columns.
We apply a combination of heuristic checks for missing values and data type errors, and a neural network based error detector that identifies attribute values not likely to co-occur in the same record.

\section{Problem Statement}
We now present the formal problem statement along with our assumptions.

\subsection{Problem Setup}
\sys takes as input a dirty training dataset $(X_{train}, Y_{train})$ where both the features $X_{train}$ and labels $Y_{train}$ may have errors, as well as a test dataset $(X_{test}, Y_{test})$ where the features may contain errors however the labels $Y_{test}$ are correct.  Although the training labels may contain errors, the test labels must be clean to ensure an unbiased measure of accuracy that is not affected by data cleaning operations.  Such labels may be collected as part of a gold standard dataset~\cite{marcus2015crowdsourced} or by cross-referencing the data with other sources~\cite{li2012truth}.  
Labels often represent directly observed phenomena such as (e.g., purchased/not purchased), while features are integrated from multiple disparate sources and subject to frequent change.
Let a record $r_i = (x_i,y_i) \in (X_{train},Y_{train})$ denote the features along with its corresponding (possibly null) label, and $r_i.y$ denote the label for the record.    Furthermore, the features may be categorical, or string-valued, in addition to numerical.

\begin{example}[Notation]
In Example~\ref{ex:lead}, the attributes $name$, $n\_emp$, $industry$, and $region$ define the schema of $X_{train,test}$, and the attribute $successful$ corresponds to the labels $Y_{train,test}$.
\end{example}

Let a classifier $C(r_i) = r_i'$ be a function that takes as input a record $r_i$ and sets $r_i.y$ to the predicted label value.
A classifier predicts $(x_i, y_i) \in (X_{test}, Y_{test})$ correctly if  $C((x_i, null)).y = y_i$.  $C$'s test accuracy is defined as the fraction of correctly predicted test records:
\[
acc(C) = \frac{|\{\forall x,y \in (X_{test}, X_{test})~:~ C((x, null)).y = y\}|}{|Y_{test}|}
\]
To generate a classifier, the user provides \textsf{train}($X_{train}, Y_{train}$) that return a classifier $C$. We model \textsf{train}($\cdot$) as a black-box and assume that the function internally performs any necessary featurization.

\begin{example}[Classification]\sloppy
The classifier $C$ can be a support vector machine predicting whether $successful = true$ based on a feature vector derived from
$name$, $n\_emp$, $industry$, and $region$.
\end{example}

\subsection{Detection and Repair Libraries}\label{s:detectorgen}
We assume that the user provides a library of detector generators $\mathcal{D} = \{d_1,\cdots\}$ and a repair library $\mathcal{F} = \{f_1,\cdots\}$.  \sys uses $\mathcal{D}$ to generate predicates that identify candidate dirty records, and selects the appropriate repair functions in $\mathcal{F}$ to those records.  

\subsubsection{Detection Generators and Predicates}
We define a predicate $p_i$ as a Boolean expression over an input record that returns the set of referenced attributes if it evaluates to $true$ and an empty set otherwise.  Based on this definition, we say that $r$ is a candidate dirty record if $p_i(r) \ne \emptyset$. For instance, $p_i(r) = r.n\_emp \le 0$ is an example of the former: if a company record contains $0$ employees, then the predicate will return $\{n\_emp\}$.  From an API perspective, we need a more expressive model than pre-defined Boolean expressions: 

First, predicate expressions may reference combinations of attributes.  For instance, if we knew that there are no oil and natural gas companies in the northwest, the predicate $p_i(r) =  (r.region == USNW \wedge r.industry \in ('OIL','NG'))$ would return $\{region, industry\}$ if such a company were detected.
Second, predicates may apply transformation functions over the input data.  For instance, the following predicate first featurizes the record using a function $g$, and applies a threshold to the first element of the feature vector: $p_i(r) = g(r)[0] > 10$.  
Third, predicate expressions may contain aggregate expressions that are computed over all records in the training dataset $X_{train}$.  For instance, the following predicate performs Quantitative Error Detection~\cite{hellerstein2008quantitative} by checking whether the record's $n\_emp$ value is further than $5$ standard deviations of the mean: 
$$p_i(r) = |r.n\_emp - avg(r.n\_emp)| > 5\times stddev(r.n\_emp)$$

To address these problems we define a detector generator $d_i$ is a function that takes the full training set as input and returns a predicate $p_i$. 
In thise sense, predicates can be derived or learned from previous data.

\subsubsection{Repair Functions}
Each repair function $f_i \in \mathcal{F}$ is a function that takes a record as input and modifies the record's attributes.  We consider two types of repairs:  {\it data repairs} are applied to the training data prior to running the training procedure, while {\it prediction repairs} modify the label of the records {\it after} the classifier makes a prediction.   

{\it Data repairs} modify the values of a training record in response to a detected error (due to a predicate).  These repair functions are free to modify the record's features, label, or simply delete the record from the training dataset.  

{\it Prediction repairs}, on the other hand, take as input the non-transformed record along with the classifier prediction, and replaces the prediction with a default value.  This is useful when the input record is too corrupted to provide a reliable prediction.  For instance, the NFL play-by-play dataset describe in Section~\ref{s:exp}, some input records contain almost all null attributes and it is more accurate to default the prediction to the most frequent label rather than attempting a repair.

Note that this section formalizes an API for these operations and subsequent sections provide one instantiation of this library.

\subsubsection{Conditional Repairs}
\sys applies repair functions to specific sets of records through the use of {\it conditional repairs}.  A conditional repair $l_k = (p_k, f_k)$ is a tuple where $p_k = d_i(X_{train}, Y_{train})$ is the output of a detector generator and $f_k \in \mathcal{F}$ is a repair function. 
A conditional repair is compiled into generation procedure that returns a repair function; the repair function takes as input a possibly cleaned record $r$, along with its original uncleaned version $r_{orig}$:
{\small\begin{verbatim}
    def generate_repair(p, f):
      def repair(r, r_orig):
        if p(r): r = f(r)   
        return r 
      return apply
\end{verbatim}}

\begin{example}[Value Canonicalization]\sloppy
The following script canonicalizes different representations for Western United States:
{\small\begin{verbatim}
    def repair(r, r_orig):
      if r.region in ('USWest', 'USWESTERN'):
        r.region = 'USW'
      return r
\end{verbatim}}
\end{example}

\vspace{0.25em}
\begin{example}[Default Prediction]\sloppy
The following script represents a conditional prediction repair that predicts $false$ if the company name is missing.  Note that the predicate is applied on the original non-cleaned record. However the classifier takes as input the cleaned version.
{\small\begin{verbatim}
    def repair(r, r_orig):
      if r_orig.name == None:
        r.y = False
        return r
      return C(r)
\end{verbatim}}
\end{example}

\noindent Finally, let $L = (l_1,\cdots,l_n)$ be a sequence of conditional data and prediction repairs that \sys generates. 
$L$ is an element in a finite universe of possible repairs denoted by $\mathcal{L}=\mathcal{D}\times\mathcal{F}$.
To apply the repairs, \sys first partitions the $L$ into two subsequences $L^d = (l_i \in L | l_i$ is data repair$)$ and $L^p = (l_i \in L | l_i$ is prediction repair$)$.  During the training phase, we apply the data repairs in sequence over the training dataset prior to training the classifier:
\begin{align}
(X'_{train}, Y'_{train}) = \{L^d(r, r) | r \in (X_{train}, Y_{train}) \}\\
C = train((X'_{train}, Y'_{train})\\
L^d(r, r) = l_k(l_{k-1}(\cdots l_1(r, r), r) r) | l_i \in L^d
\end{align}

Finally, \sys constructs the final classifier $C_{L}$ by combining the prediction repairs $L^p$ with the trained classifier $C$.  It first identifies the last prediction repair $l^* \in L^p$ whose predicate matches the test record.  
$$l^* = \argmax_{l_i \in L^p \wedge l_i(r) = true} i$$
If no such prediction repair is found, \sys  returns the classifier prediction on the cleaned record, otherwise it applies $l^*$:
$$C_{L}(r) = \begin{cases}
    C(L^d(r, r))& \text{if } l^*\textrm{\ not\ found}\\
    l^*(L^d(r, r), r) & \text{otherwise}
\end{cases}$$

\subsection{Scope and Assumptions}
As a class of errors, we focus on domain integrity constraints, i.e., a set of allowed values in each attribute's domain--an error being defined as an attribute value not in this set.
Given a violation, we assume that each of the repair actions sets the attribute to an allowed value.
This assumption avoids a fixed-point iteration, also called the ``chase algorithm''~\cite{aho1979theory}, which repairs that cause additional errors.
This greatly simplifies the specification of $\mathcal{L}$ the set of possible data cleaning operations--in our experiments, $|\mathcal{L}|$ varied from $192$ to $1076$.
Next, we assume that each record in a relation corresponds to a single example (features and labels), and the analyst wants to learn a classifier that predicts labels from features.
Finally, we assume that the labels of the test data are clean since \sys relies on uncorrupted labels to estimate the model's accuracy.

\subsection{Problem Statement}
Given these assumptions, we define the repair selection problem:

\begin{problem}[\sys Repair Selection]\sloppy
Given $(X_{train}, Y_{train})$, $(X_{test}, Y_{test})$, a library of detector generators $\mathcal{D}$ and of repair functions $\mathcal{F}$, and a training procedure $train$, identify the optimal sequence $L^*$ of $B$ conditional repairs such that the resulting classifier $C_{L^*}$  maximizes prediction accuracy on $(X_{test}, Y_{test})$:
$$L^* = \argmax_{L \in \mathcal{D}\times\mathcal{F}} acc(C_L)$$
\end{problem}

Greedy solutions that select the top $B$ individual condition repairs will often fail since they might select highly correlated repairs (e.g., imputing a missing value with the mean, and the median).
Instead, it is desirable for an approach to take the mispredictions from previous conditional repairs into account.  This is the reason we applied a boosting-based approach towards selecting conditional repairs, described in the next section~\cite{schapire2003boosting}.

\begin{figure}\centering
\includegraphics[width=\columnwidth]{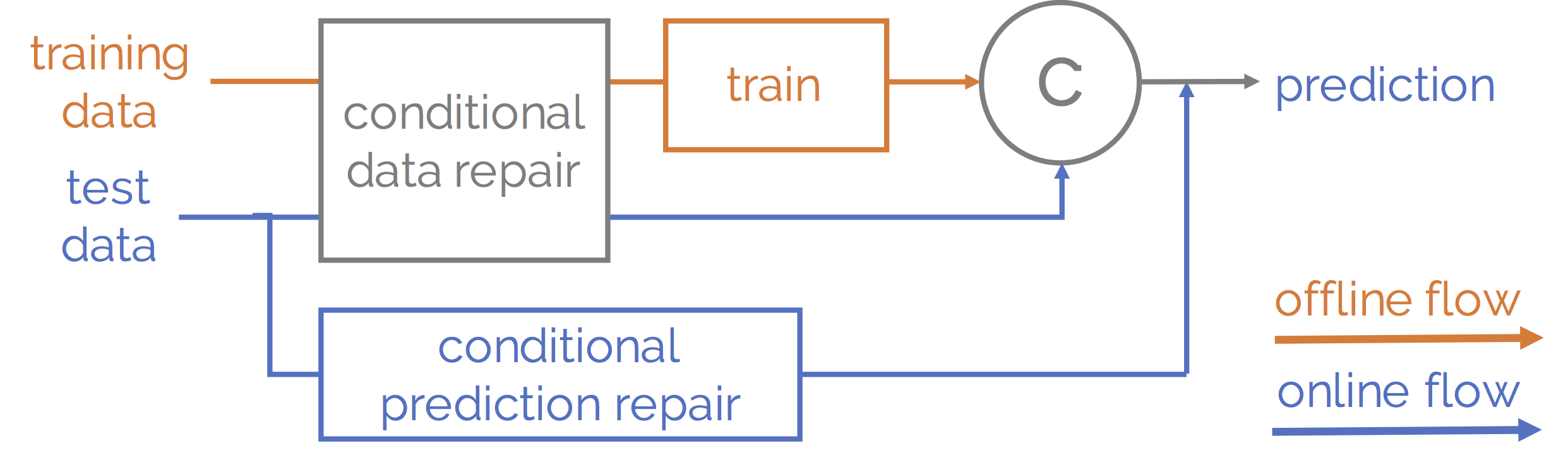}
\caption{Offline (\orange{orange}) and online (\blue{blue}) workflows.}
\label{fig:workflow}
\end{figure}

Figure~\ref{fig:workflow} summarizes the training and prediction workflows given the optimal sequence of conditional repairs $\mathcal{L}^*$.  The \orange{orange line} depicts the training process, which first applies the conditional data repairs to the training dataset, and calls \texttt{train()} to generate classifier $C$.  The \blue{blue lines} depict how \sys generates a prediction for a test record: the classifier $C$ makes a prediction using the record cleaned by the conditional data repairs.  In addition, the conditional prediction repair checks the uncleaned test record to decide whether to return the classifier prediction or a default value.

\section{Repair Selection Algorithm}
The key insight of this paper is that the problem described in Section 3.3 can be addressed with statistical boosting.

\subsection{Overview of Boosting}
Ensemble methods construct predictions from combinations of predictors.
Boosting, a type of ensembling, is based on the observation that finding many ``weak learners'' is often significantly easier than finding a single, highly accurate predictor. 
The boosting algorithm calls this ``weak'' or ``base'' learning algorithm repeatedly feeding it a  weighting over the training examples.
Each time it is called, the base learning algorithm generates a new weak prediction rule, and after many rounds, the boosting algorithm must combine these weak learners into a single prediction rule that, hopefully, will be much more accurate than any one of the weak learners.

We will first introduce the classical AdaBoost algorithm for binary classifiers.
This is without a loss of generality since we can use an all-versus-one technique to handle multi-class classification.
The algorithm takes as input a training set of features and labels $(X,Y)$--assume that the labels are $\{-1, 1\}$.
AdaBoost calls a given weak learner repeatedly in a series of rounds. 
The algorithm re-weights the dataset after each round.
By training on a weighted dataset, we mean that it finds a learning from a set of permissible learners that maxmimizes the weighted accuracy.
For weighting function $W(x,y) \mapsto \mathbb{R}_+$:
\[
acc(C, W) = \frac{\sum_{x,y} W(x,y) \mathbf{1}(C((x, null)).y = y)}{\sum_{x,y} W(x,y) }
\]
Initially, all weights are set equally, but on each round, the weights of incorrectly classified examples are increased so that the learner is forced to focus on the hard examples in the training set.

Formally, the AdaBoost algorithm~\cite{freund1995desicion} proceeds as follows:
\begin{algorithm}
\KwData{(X, Y), $\alpha$}
Initialize $W^{(1)}_i = \frac{1}{N}$\\
\For{$t \in [1, T]$}{
  $C_t$ = Train weak learner on dataset weighed by $W^{t}_i$ \\
  $\epsilon_t$ = Calculate weighted classification error \\
  $\alpha_t = \ln(\frac{1-\epsilon_t}{\epsilon_t})$ \\
  $W^{(t+1)}_i \propto W^{(t)}_i e^{-\alpha_t y_i C_t(x_i)}$: down-weight correct predictions, up-weight incorrectly predictions.
}
\Return $C(x) = \text{sign}(\sum_t^T \alpha_t C_t(x) )$
\caption{AdaBoost Algorithm}
\label{alg:adaboost}
\end{algorithm}

To be able to understand Adaboost theoretically, we require an assumption called the \emph{weak learning assumption}. That is, we assume that our weak learner can consistently find classifiers which classify the data correctly at better than random guessing for any weighting of the dataset. 

\subsection{Why Boosting?}
The key difference from naive feature selection algorithms is that it selects over the space of models rather than the space of features.
If we have repair operations that cannot simply be represented as columnar operations (e.g., removing a record), this is a preferred solution.
Similarly, it makes few assumptions about how the attributes are aggregated into a model.

In our problem, each of the library elements will define a weak learner.
Given the dataset $R$, we can apply $L \in \mathcal{L}$ and then train the classifier to return $C_L$. 
The weak learners are evaluated on the clean test labels, which dictates weighting.
Modeling the selecting process as a statistical boosting allows us to make relatively few assumptions about the classifier and the data cleaning operations. 
Instead of having to reason about composing different data cleaning operations (and how compositions may affect accuracy), we are reasoning about a weighted consensus of classifiers trained with different data cleaning approaches.

\subsection{Boost-and-Clean Algorithm}\label{s:boostalg}
The boosting algorithm weights the dataset depending on mispredictions, focusing future effort on the ensembles current mispredictions.
In each round, we find the $L \in \mathcal{L}$ that generates the classifier with highest test accuracy on the weighted data.
After selection, we recalculate the wights.
Repeat until $B$ cleaning operations are selected, by selecting the operation that performs best with updated weights.
The result is a new classifier $C_{clean}$ that is derived from the ensemble.
As before, without loss of generality we present the binary classification case with labels in $\{-1,1\}$.

\begin{algorithm}
\KwData{(X, Y)}
Initialize $W^{(1)}_i = \frac{1}{N}$\\

$\mathcal{L}$ generates a set of classifiers $\mathcal{C} \{C^{(0)}, C^{(1)},...,C^{(k)}\}$ where $C^{(0)}$ is the base classifier and $C^{(1)},...,C^{(k)}$ are derived from the cleaning operations.\\

\For{$t \in [1, T]$}{
  $C_t$ = Find $C_t \in \mathcal{C}$ that maximizes the weighted accuracy on the test set. 
  $\epsilon_t$ = Calculate weighted classification error on the test set
  $\alpha_t = \ln(\frac{1-\epsilon_t}{\epsilon_t})$ 
  $W^{(t+1)}_i \propto W^{(t)}_i e^{-\alpha_t y_i C_t(x_i)}$: down-weight correct predictions, up-weight incorrectly predictions.
}
\Return $C(x) = \text{sign}(\sum_t^T \alpha_t C_t(x) )$
\caption{Boost-and-Clean Algorithm}
\label{alg:rsa}
\end{algorithm}

The algorithm has a few intuitive properties: (1) it prioritizes cleaning operations that improve performance, (2) if no such operations exist it does no worse than the base classifier, and (3) it is agnostic to the implementation of the classifiers.
The basic runtime of the algorithm is polynomial in both the number of cleaning operations and size of the dataset. In the next subsection, we will describe optimizations.

\begin{proposition}[Time Complexity]
The time complexity of Boost-and-Clean is $\mathbf{O}(k^2 N_{test} + k N_{train})$, where $k$ is the number of data cleaning operations, $N_{test}$ is the number of test tuples, and $N_{train}$ is the number of training tuples.
\end{proposition}

Boosting is well-understood statistically, and we can further bound the error on our clean test set (follows from~\cite{schapire2003boosting}). This requires the application of the weak learning assumption which means that for any weighting of the test set, we can find one library component that classifies better than random guessing. This assumption is a theoretical assumption needed for the formal guarantee. Even if this assumption does not hold, \sys will still identify $B$ cleaning operations.

\begin{proposition}[Error Bound]
Assuming that the weak learning assumption holds, for a budget of $B$ cleaning operations, the error rate of Boost-and-Clean on the test dataset decreases as $\mathbf{O}(e^{-2B})$.
\end{proposition}

\section{The \sys System}\label{s:arch}
 
\begin{figure}[t]
\centering
\includegraphics[width=\columnwidth]{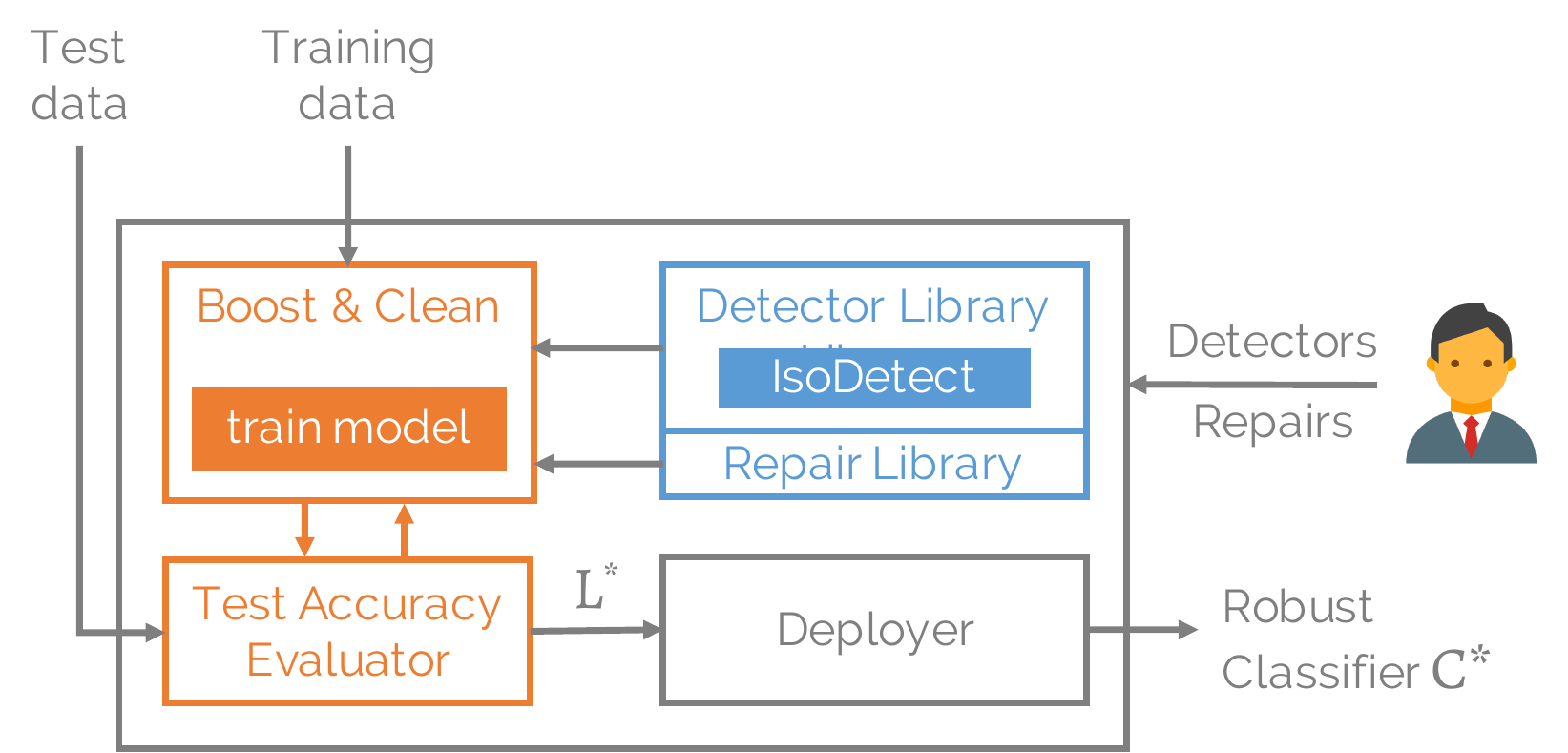}
\caption{\sys system architecture.}
\label{f:arch}
\end{figure}

Figure~\ref{f:arch} depicts the primary components of the system architecture. The \blue{blue} component manages the detection and repair libraries while the \orange{orange} components execute the Boost-and-Clean algorithm to generate the sequence of conditional repairs $\mathcal{L}^*$.  To generate $\mathcal{L}^*$, \sys takes as input training and test datasets, where the only restriction is that the test labels are correct. 

\sys first executes the library of detector generators on the training dataset to produce a set of candidate predicates (Section~\ref{s:detectorgen}).    The {\it Boost-and-Clean} component takes these predicates, along with the repair functions, as input and selects a subset of the pairs (conditional repairs).  To do so, it iteratively selects the next conditional repair by appending it to the sequence of repairs so far, training a new classifier using the sequence and evaluating it using the {\it Test Accuracy Evaluator}.  We also call this process {\it repair selection}.  Finally, the conditional repairs $\mathcal{L}^*$ are sent to the {\it Deployer}, which compiles the sequence into a  classifier that can detect and repair errors in the test records so that the predictions are robust to the detected data errors.  

\sys is pre-populated with a library of detector generators and repair functions that work well in practice (Section~\ref{s:exp}).  However,  developers can also specify custom detection or repair functions that fit their domain.  To further simplify how the system can be used, developers can implement familiar feature extraction functions, and the {\it IsoDetect} component will automatically translate them into detector generator functions.

\subsection{Detectors}
The ability for a data cleaning system to accurately identify data errors relies on the availability of a set of high-quality error detection rules~\cite{DBLP:conf/sigmod/ChuIKW16}.  To help developers easily implement new detector generators, we have implemented \detectlib, a library that transforms feature extraction functions into detector generators automatically.  Also, to help developers quickly bootstrap their data cleaning process, \sys includes a pre-populated library of simple feature extraction functions that are effective at detecting data errors across a wide range of domains and data science applications (Section~\ref{s:exp}).  We note that these libraries are not meant to replace domain knowledge, but rather to address routine problems.

\subsubsection{The \detectlib Library}
Although developers may directly implement detector generators as described in Section~\ref{s:detectorgen}, we found that many cleaning detection algorithms in machine learning workflows, including \company, follow a fixed outlier-detection structure: first they convert a record into a feature vector,  select threshold(s) over one or more features in the vector such that if the thresholds are exceeded, then the record is identified as a candidate dirty record.  Although developers often have a clear sense of useful features to extract from the dataset, it is often unclear how to tune the thresholds and for which features thresholds should be defined for.  

For this reason, \detectlib employs a modular approach that automatically performs the latter threshold tuning task so that developers can focus on feature engineering.   Developers simply define featurization functions that map records into a numerical feature vector; for each featurizer, \detectlib executes it and performs outlier detection to select the combination of features and thresholds that will identify the outliers.  

We considered a number of outlier detection algorithms to use.  A recently popular approach is a robust estimator of the sample population variance called Minimum Covariance Determinant~\cite{rousseeuw2011robust} (MCD), which has been used in systems such as MacroBase~\cite{bailis2016macrobase}.  The key limitation is that the technique is computationally expensive and can have degenerate results (due to rank-deficient covariance matrices).

We instead use a variant of Random Forest classification, called Isolation Forests~\cite{liu2008isolation}.  The Isolation Forest is inspired by the observation that outliers are more easily separable from the rest of the dataset than non-outliers.  It grows a forest of isolation trees, where each tree is randomly grown---it selects a random attribute and a random threshold value---until a leaf node contains a single record.  The length of the path to the leaf node is a measure for the outlier-ness of the record---a shorter path more strongly suggests that the record is an outlier.  The isolation forest creates a large set of isolation trees and classifies records with short average path length as outliers.  In contrast to computational expensive algorithms such as MCD, Isolation Forests have a linear time complexity and very small memory requirements.  

A nice property of the technique is that the resulting forest can be simplified and efficiently compiled into simple threshold rules. 
For example, if a featurizer extracts a scalar feature in $\mathbb{R}$ (e.g., $n\_emp$), the Isolation Forest will generate a single-attribute threshold rule (e.g., $n\_emp > 2$).  
 We experimented with alternative outlier detection techniques (e.g., Minimum Covariance Determinant) but found that the Isolation Forest provided the best trade-off between runtime and accuracy.
 
\stitle{Hyper-parameters: } The Isolation Forest has a few hyper-parameters that need to be tuned, namely, the maximum branch length and a threshold that determines outlier v.s. not outlier. As much as possible, we have tuned these hyper-parameters generically with sensible defaults. Experimentally, we find that performance does not suffer too much with the default parameters.
In one of our experiments, we used a single hyper-parameter setting across all datasets and evaluate accuracy. 
 
\stitle{False Positive and False Negatives: } In general, a predicate learned from data will have false positives and false negatives.
This is why the boosting selection is important.
If a predicate is too uncertain and applies the repair to many spurious records, then it will reduce classification accuracy.
By boosting, we can protect the system against uncertain detectors.

\subsubsection{Pre-populated Featurizers}
\noindent We have implemented four classes of featurizers:

\stitle{Numerical Attributes: } This featurizer projects the numerical attributes into a feature vector.  We find that this is effective at identifying numerical outliers that are statistically different from the rest of the data. For example, the sensor dataset has sensor readings on the order of $300C$ when typical readings are $17C$ (Section \ref{exp:comp}).

\stitle{Missing Values: }  We manually enumerated a set of regular expression patterns that commonly describe missing values in a database.  These include \textsf{NULL} attribute values; patterns such as empty string, NaN, Inf; or values whose string representations lack alphanumeric characters. Each pattern corresponds to a boolean feature in the extracted feature vector.   Although this is a hand-curated list, it is easily extensible and is effective at identifying missing values. 11 out of the 13 experimental datasets, had some form of missing value error.

\stitle{Parsing/Type Errors: } We use each attribute's type signature to check whether an attribute value matches the type signature.  For numerical attributes, the featurizer outputs whether the entry can be parsed into a floating point number or an integer. For dates and address types, we check whether common components (e.g., Month, Day, Year for dates, Street, City for addresses) are found using common date and address parsing libraries.   This means that the entry has a minimum of the required components (Month, Day, Year) or (Street, City, State).  This is implemented with standard python libraries: \textsf{usaddress} and \textsf{datetime}. The FEC dataset had a small number of rows with the wrong number of columns, leading to data type errors. 
We have a ``short-circuit'' routine for featurizers to declare an attribute an instant failure without passing it into \detectlib.

\stitle{Text Errors using Word Embeddings: }
Although the above featurizers are effective for quantitative attributes, many datasets contain string-valued and categorial attributes that are not amenable to the above approaches.  Also, naively featurizing text attributes using, e.g., hot-one encoding, can easily increase the dimensionality of the dataset to tens of thousands or millions of dimensions---even a two-attribute relation with one numerical and one string-valued attribute, may have thousands of features.  The statistical power of outlier detection techniques rapidly diminishes in the high-dimensional feature-spaces.

In response, we borrow the recent concept of text embeddings from Natural Language Processing to featurize record values into a lower-dimensional vector-space.  
Text embeddings are models trained on a corpus of documents that embed words from the corpus into a vector-space where nearby vectors are similar words.
This allows one to featurize string and categorical values into numerical vectors, and evaluate similarity relationships between documents in this vector-space.

We adapt the popular \textsf{word2vec} model~\cite{mikolov2013distributed} to structured outlier detection. 
We treat each record as a document, where each attribute value is a ``word''. 
The model learns to embed attributes that co-occur in the same records closer in the vector-space.
Thus, each attribute value is mapped to a vector, and each record is the concatenation of its attributes' vectors.  
The isolation forest then takes this vector as input to generate an appropriate predicate.  

The isolation forest has the crucial property that the anomaly detection criteria are axis-aligned cuts.  Since each set of features corresponds to a record attribute, we can directly translate threshold violations into the data attributes that are erroneous.
In our experiments, we find that this approach is effective at detecting a variety of categorical errors that we did not explicitly code for.
One common example is when a header record (i.e., one specifying the names all the attributes) is included in the dataset.
The model identifies this record as containing values not typically present in the dataset.
Similarly, this module detected oddly formatted codes in the FEC dataset.

\subsubsection{Adding Custom Featurizers} 
Our featurizers are meant to be a starting point that is supplemented by domain specific modules. For example, if the data scientist knows that employee ids in a database must match a particular pattern, she can build featurizers to parse these ids. Similarly, if the data scientist knows something about the structure of the features (e.g., they are time-series), she can add in other features such as frequency components derived from an FFT. All of these featurizers must return a vector and a mapping between vector components and base-data attributes.

\subsection{Repairs}
In addition to detector generators, \sys is pre-populated with a set of simple repair functions.  A function is applied to all records identified by a detector's predicate.  In the following five repair functions, the first three can be used as data and prediction repair functions,  whereas the fourth is for data repair, and the last is for prediction repair:

\stitle{Mean Imputation (data and prediction): } Impute a cell in violation with the mean value of the attribute calculated over the training data excluding violated cells.

\stitle{Median Imputation (data and prediction): } Impute a cell in violation with the median value of the attribute calculated over the training data excluding violated cells.

\stitle{Mode Imputation (data and prediction): } Impute a cell in violation with the most frequent value of the attribute calculated over the training data excluding violated cells.  In contrast to mean and median imputation, this is also applicable to non-numerical attributes.

\stitle{Discard Record (data): } Discard a dirty record from the training dataset.  This restriction to only training data is to ensure that a degenerate solution---simply deleting all test data---is disallowed. 

\stitle{Default Prediction (prediction): } Automatically predict the most popular label from the training dataset for a row that matches the conditional repair's predicate.


\subsection{Optimizations}
In order to borrow the error bound guarantees, the Boost-and-Clean algorithm (Algorithm~\ref{s:boostalg}) is a direct translation of AdaBoost and a naive implementation of the algorithm can be expensive.  Namely, it takes as input the cross product of predicates and repair function and, in each iteration, requires applying the conditional repairs, and training and testing a classifier for each conditional repair.  
To reduce the cost, we employ optimizations to address three bottlenecks in the naive algorithm.  These optimizations are developed specifically to speed up the Boost-and-Clean repair selection process, and we did not attempt to optimize the detection nor data parsing and loading steps of the end-to-end system.  For the purposes of this paper, we additionally assume that the test datasets fit into memory (the training data do not have this restriction).

\stitle{Prediction Materialization: } Our first observation is that the cleaning operations actions do not change between iterations of the boosting algorithm--only the weights for computing the accuracy change.  Thus, we pre-train the classifiers $C_i$ corresponding to each conditional repair $l_i$, and materialize their predictions on the test records $C_i(X_{test})$.  

\stitle{Prediction Indexing: } Computing the score for each classifier requires retrieving the the weights of the test records that are mis-predicted and correctly predicted.  We speed up these lookups using a hash index for each classifier that that maps prediction labels to their corresponding test records.

\stitle{Parallelization: }  Finally, many of the operations in Boost-and-Clean, such as classifier training and scoring, must be performed for each conditional repair.  We create and execute a thread for each conditional repair to perform each task in parallel.  We leave more advanced optimization techniques such as within-training parallelization~\cite{recht2011hogwild} and sampling the training data for future work.
\section{Experiments}\label{s:exp}
In this section, we present the results of our experiments.  We execute \sys on $12$ datasets based on three sets of real-world cases---machine learning competitions, data analysis pipelines, and \company---and report accuracy measures and end-to-end runtime.  Then, we present a series of micro-benchmarks that evaluate each of the modules of \sys.  Our goal is to understand the conditions where automated cleaning is able to accurately detect and repair data in a way that improves the held-out test accuracy. 

In particular, we evaluate three hypotheses: 1) Compared to baselines, the cleaning operations selected by \sys result in a greater improvement to the downstream classification accuracy; 2) \sys automatically detects a large fraction (in comparison to hand-written rules) of the errors across several different datasets; and 3) The optimizations that we design for \sys allow us to run at a reasonable wall clock time.

\subsubsection{Setup}

To the best of our knowledge, there does not exist a comparable general purpose ML+Data Cleaning system to \sys in industry or academia.  We evaluate \sys against a number of baseline approaches inspired by solutions proposed in literature.   These baselines are described in the subsequent experimental subsections.
We used the following setup for our experiments. 

\stitle{Test Data: } For each of the datasets, we defined a 20\% held-out test dataset. We assumed that the labels in this test dataset were clean, as per the assumption in \sys. To avoid overfitting, we carefully designed the accuracy evaluation experiments for \sys by using a ``doubly'' held out test dataset: the test dataset used to optimize \sys is different from a completely unseen 20\% test dataset that is solely used to report the final prediction accuracy.
Training is performed on the remaining 60\% of the data.

\stitle{Models: } We used the \textsf{sklearn} Random Forest classifier.  The training procedure uses a set of standard featurizers (hot-one encoding for categorical data, bag-of-words for string data, numerical data as is) in a similar fashion as~\cite{gokhale2014corleone}.  Note that these featurizers are used as part of the black-box training procedure and are distinct from those used in the detector generator library. 
We describe hyper-parameter settings for each technique in the text of each experiment.
As much as possible, we attempted to use the library default parameters.

\stitle{Timing: } In all of our experiments, we used standard classification models and featurization techniques from Python \textsf{sklearn}.
The classifiers were trained in Python 2.7 and timing experiments were run on an Amazon EC2 m4.16xlarge instance\footnote{64 virtual cpus and 256 GiB memory}.

\begin{table*}[t]
\centering
\begin{tabular}{|l|r|r|r|r|r|r|r|r|r|r|r|}
\hline
ML Competition& \#rows & \#cols & NC & Q &	IC & Q+IC &	Best-1 &	Worst-1 &	BC-3 & BC5 & Rel. Improvement\\
\hline
USCensus	&32561&15&0.85&	0.82&	0.86&	0.84&	0.87&	0.79&	0.88&	\pop{0.91} & +4.5\% \\
Emergency &11176&9&	0.67&	0.72&	0.67&	0.72&	0.72&	0.66&	0.72&	\pop{0.75} & +4.7\%\\
Sensor	&928991&5&0.92&	0.93&	0.92&	0.89&	0.92&	0.8&	\pop{0.94}&	0.94 & +1.3\%\\
NFL	&46129&65&0.74&	0.74&	0.76&	0.75&	0.76&	0.74&	0.79&	\pop{0.82}& +5.1\%\\
EEG	&2406&32&0.79&	0.82&	0.79&	0.83&	0.83&	0.7&	0.85&	\pop{0.89}& +6.8\%\\
Titanic	&891&12&0.83&	0.72&	0.83&	0.76&	0.83&	0.69&	0.83&	\pop{0.84}& +1.1\%\\
Housing	&1460&81&0.73&	0.76&	0.73&	0.77&	0.77&	0.65&	\pop{0.81}&	0.76& +5.1\% \\
Retail	&541909&8&0.88&	0.88&	0.91&	0.91&	0.91&	0.87&	0.94&	\pop{0.95}& +4.3\% \\
\hline
\hline
Data Analytics &\#rows & \#cols & NC & Q &	IC & Q+IC &	Best &	Worst &	BC-3 & BC5 & Rel. Improvement\\
\hline
FEC  & 6410678 & 18 & 0.62 & 0.53 & 0.61 & 0.57 & 0.71 & 0.51 & 0.74 & \pop{0.77} &  +8.4\% \\
Restaurant (Multiclass) &758&4& 0.42 & 0.42 & 0.58 & 0.68 & \pop{0.62} & 0.36 & 0.61 & 0.60 & (1.61)\% \\
\hline
\hline
Company X &\#rows & \#cols & NC & Q &	IC & Q+IC &	Best &	Worst &	BC-3 & BC5 & Rel. Improvement\\
\hline
Dataset 1 (AUC) &76684&6&0.60&0.60&0.60&0.60&0.61&0.59&0.66&\pop{0.69}& +13.3\% \\
Dataset 2 (AUC) &83986&6&0.55&0.55&0.52&0.55&0.55&0.52&0.61&\pop{0.63}& +14.5\%\\
\hline
\end{tabular}
\caption{End-to-end accuracy results for each dataset and experimental method. We report standard classification accuracy.  The right column summarizes the absolute accuracy improvement over the best non BC-3/5 approach.  The \company datasets have high class imbalances cause artificially high accuracy statistics,  so we report AUC statistics for those datasets instead.}
\label{tab:accuracy}
\end{table*}

\subsection{End-to-End Accuracy}
In our first experiment, we evaluated the accuracy of \sys compared to the baselines.
We tried to minimize hyper-parameter tuning as much as possible to simulate a real-scenario where extensive tuning and parameter search might be expensive.

\subsubsection{Methods}

\stitle{No Cleaning (NC): } We train a model without any modification to the training or test data.

\stitle{Quantitative (Q): } We train a model where only the isolation forest over the numerical attributes is used to detect errors.
Errors in both training and test are imputed with a mean value.

\stitle{Integrity Constraint (IC): } We read through each dataset to identify a set of anomalous values for each non-numerical attribute on a best-effort basis.  We then codified these as integrity constraint rules, and corrected the identified errors using a statistical distortion minimization metric as in~\cite{prokoshyna2015combining}. Statistical distortion minimizes the statistical distance to some ideal distribution (e.g., Power Law or Gaussian). We set these distributions manually by inspecting the data when possible.

\stitle{Quantitative + IC (Q+IC): } We use both the quantitative and integrity constraints for detection. For repair, we apply an imputation with a default value. For categorical and string-valued attributes, this the most frequent value. For numerical attributes, this is the mean value.

\stitle{Best Single (Best-1): } We run \sys with $B=1$ and identify the  single best conditional repair.

\stitle{Worst Single (Worst-1): } We run \sys with $B=1$ and identify the single worst conditional repair..

\stitle{BC-3: } We run \sys with $B=3$.

\stitle{BC-5: } We run \sys with $B=5$.

\subsubsection{ML Competition Datasets}\label{exp:comp}
We downloaded 8 binary classification datasets from Kaggle competitions and benchmarks in the UCI ML repository.  
These datasets have been extracted, structured, and published.
Nevertheless, they contain missing values, numerical outliers, and pattern errors (oddly formatted values).
For this set of experiments, we used a single hyper-parameter setting for all the detectors and classification models (default \textsf{sklearn} library setting). 
We briefly describe each dataset and their errors below:

\vspace{0.5em}\noindent\textbf{USCensus: } This dataset contains US Census records for adults and the goal is to predict  whether the adult earns more than $50,000$ dollars. It contains 32,561 records with 15 numerical and categorical attributes. This dataset contained missing values and coding inconsistencies.

\vspace{0.5em}\noindent\textbf{NFL: } This dataset contains play-by-play logs from US Football games. The dataset contains 46,129 records with 65 numerical, categorical, and string-valued attributes. Given the record, the classification objective is to determine whether the next play the team runs is a run or a pass play.
The dataset contains a significant number of missing values and ``sentinel'' records that mark the end of a log sequence. The sentinel records do not signify a play but rather signify a timeout, end of quarter, or end of the game.

\vspace{0.5em}\noindent\textbf{EEG: } This is a dataset of EEG recordings. 
The training data is organized into ten minute EEG clips labeled "Preictal" for pre-seizure data segments, or "Interictal" for non-seizure data segments. 
There are 2406 records each of which is a variable-length time-series of 16 attributes. We featurize this dataset into records of 32 attributes--the mean and variance over the length of the time-series. 
This dataset primarily contains numerical outliers, the clips have spurious readings.

\vspace{0.5em}\noindent\textbf{Emergency: } This dataset contains records on 911 calls from Pennsylvania. There are 111,766 records with 9 attributes. Given the record, the classification challenge is to determine whether the emergency service response time will be less than $5 min$. This dataset contains missing values, and spurious locations not served by the 911 center.

\vspace{0.5em}\noindent\textbf{Sensor: } The Intel sensor dataset~\cite{data, DBLP:journals/pvldb/0002M13, wang1999sample} contains 928,991 temperature, humidity, and light sensor readings a sensor deployment. The classification task is to predict whether the readings came from a particular sensor (sensor 49). This dataset primarily has numerical outliers.

\vspace{0.5em}\noindent\textbf{Titanic: } This dataset contains 891 records from the Titanic manifest with 12 attributes. The classification objective is to determine whether the passenger survived or not. There are missing values and string formatting errors.

\vspace{0.5em}\noindent\textbf{Housing: } The housing dataset contains 1460 records and 81 attributes of house price listings. The classification objective is to determine whether the listed house will be sold above 750000. 
This dataset contains missing values as well as numerical outliers.

\vspace{0.5em}\noindent\textbf{Retail: } The online retail dataset contains 541,909 records of online retail purchases with 8 attributes. The classification objective is to predict whether the purchase occurred in the United Kingdom.
This dataset contains numerical errors where some purchased quantities are reported as negative.

\vspace{1em}
The first set of rows in Table \ref{tab:accuracy} present the predictive accuracy of models trained with \sys on the completely unseen test data.  In all experiments, the model trained with one of the \sys  approaches was the most accurate.
The quantitative baseline performed well when the errors were clear numerical outliers (e.g., Sensor and  EEG).  However, its performance suffered in datasets with missing values or formatting errors, and {\it degraded model accuracy} in the US Census dataset.
Conversely, the integrity constraint approach worked well for non-numerical errors, however it was not useful for Emergency, EEG, Housing, nor Sensor.
The naive union of (Q+IC) has difficulty composing the two operations in the US Census dataset and degrades accuracy as compared to quantitative or integrity constraint alone in several datasets.  Finally we compare and find up to a 14\% difference between the best and worst repairs when using \sys.  These results emphasize the need for an automatic search solution that can avoid repairs that are ineffective or reduce accuracy.

BC-3 and BC-5 improve the predictive performance of the models.
In all of the datasets, we found that either BC-3 or BC-5 had the highest test accuracy.
There is an interesting reason why BC-3 is more accurate than BC-5 in two cases.
Consider the case where there are only three types of errors in the dataset.
Then BC-3 would in principle select cleaners to address them. The remaining two cleaners would just add noise.
Our evidence suggests that this happened in the two datasets where the errors were mostly concentrated on a handful of attributes.

\subsubsection{Data Analytics}
The next class of datasets that we considered were datasets known to have significant errors--unlike the relatively clean competition datasets. These are two datasets that were used in previous data cleaning papers, and we designed classification tasks based on the datasets.
Unlike the ML competition datasets, we tuned the classifier and detector hyperparamters for each dataset. 
The accuracy results are presented in the second set of rows in Table \ref{tab:accuracy}.

\vspace{0.5em}\noindent\textbf{Federal Election Commission Contributions: } The FEC provides a dataset of election contributions of 6,410,678 records with 18 numerical, categorical and string valued attributes. This dataset has a number of errors. There are missing values, formatting issues (where records have the wrong number of fields causing misaligment in parsing), and numerical outliers (negative contributions).

Our classification objective was to determine whether the contribution would be above or below $100$ dollars. Due to the severity of the errors in the dataset, there is nearly a 15\% difference between the prediction accuracy of a classifier with and without \sys.
Furthermore, a purely quantitative approach is not useful for this dataset.
An integrity constraint based method improves accuracy but the automatic imputations are unreliable on this data.
Furthermore, it is difficult to express a problem like row misalignment as a integrity constraint.

We find empirically that the alignment is better detected by the word2vec error detector in \sys.
As a result the best single cleaner is using the word2vec error detector.
This is improved by combining this with quantitative checks for numerical outliers and missing values.
In all, \sys with a budget of 5 improves accuracy 8.4\% over the best single cleaner. 

\vspace{0.5em}\noindent\textbf{Restaurant Dataset: } The restaurant dataset has 758 distinct records and 4 attributes. This dataset has typically been used as a benchmark for entity resolution since records are duplicated with minor inconsistencies.
We designed a multi-class classification task to see if we could predict the city from record.
One of the major inconsistencies was additional attributes appended to the restaurant category.

On this dataset, we see a negative result from \sys. Our test error decreases as we increase the number of selected cleaners. We speculate this is due to overfitting due to the extremely small size of the dataset  ($<1000 records$) combined with the expressiveness of the classifiers model.

\subsubsection{Company X Experiments}
We applied \sys to two datasets from Company X of 76,684 records and 83,986 records respectively (each with 6 attributes). 
All of these attributes were inferred as categorical by our type inference module.
What made these datasets interesting was a significant class imbalance, where most records were labeled $0$ and few were labeled $1$.
Because of this imbalance, the accuracy of simply predicting the common label performs nearly perfectly, and we instead report the AUC classification score.  Furthermore, due to data confidentially, we were only able to acquire aggregate statistics about the data cleaning results.

In these two datasets, the primary errors were detected by the missing values and word2vec featurizers. 
In Dataset 1, over 50\% (40,164) of the rows contain some type of error (either a missing values or anomalous categorical value in at least one attribute).
Similarly, over $95\%$ of Dataset 2 (80,168  rows) contained at least one instance of missing value or anomalous categorical errors.  The detailed AUC results are reported in the last set of rows in  Table \ref{tab:accuracy}.
\sys with 5 selected cleaners achieves an absolute improvement of 8\% ($14\%$ relative improvement) in both datasets over the next best non-\sys alternative, and a slightly larger improvement over not cleaning the dataset at all.
Interestingly, we found that the integrity constraint approach (IC) reduced the AUC results in dataset 2.

\begin{figure}[t]
\centering
\includegraphics[width=0.8\columnwidth]{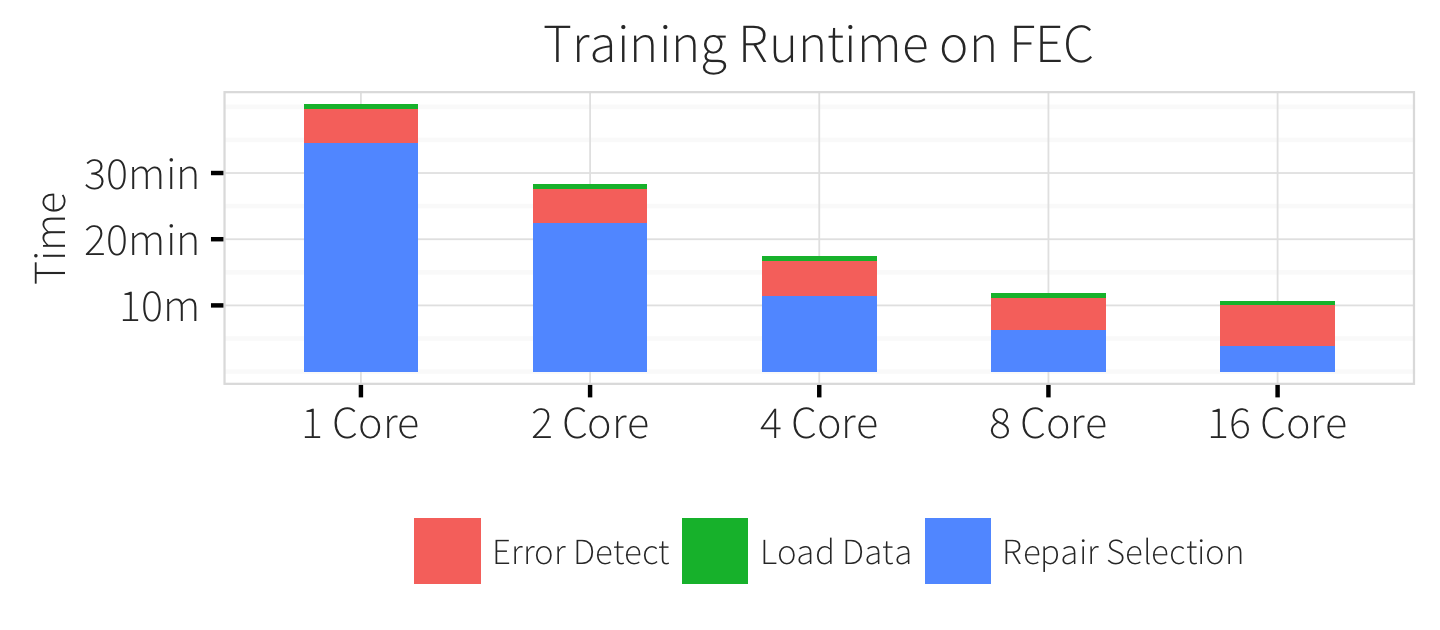}
\caption{Training runtime on a 6M record dataset (1.5GB). The repair selection scales due to the parallelization optimization, however we did not parallelize the other steps.\label{exp:runtime}}
\end{figure}

\begin{figure}[t]
\centering
\includegraphics[width=0.8\columnwidth]{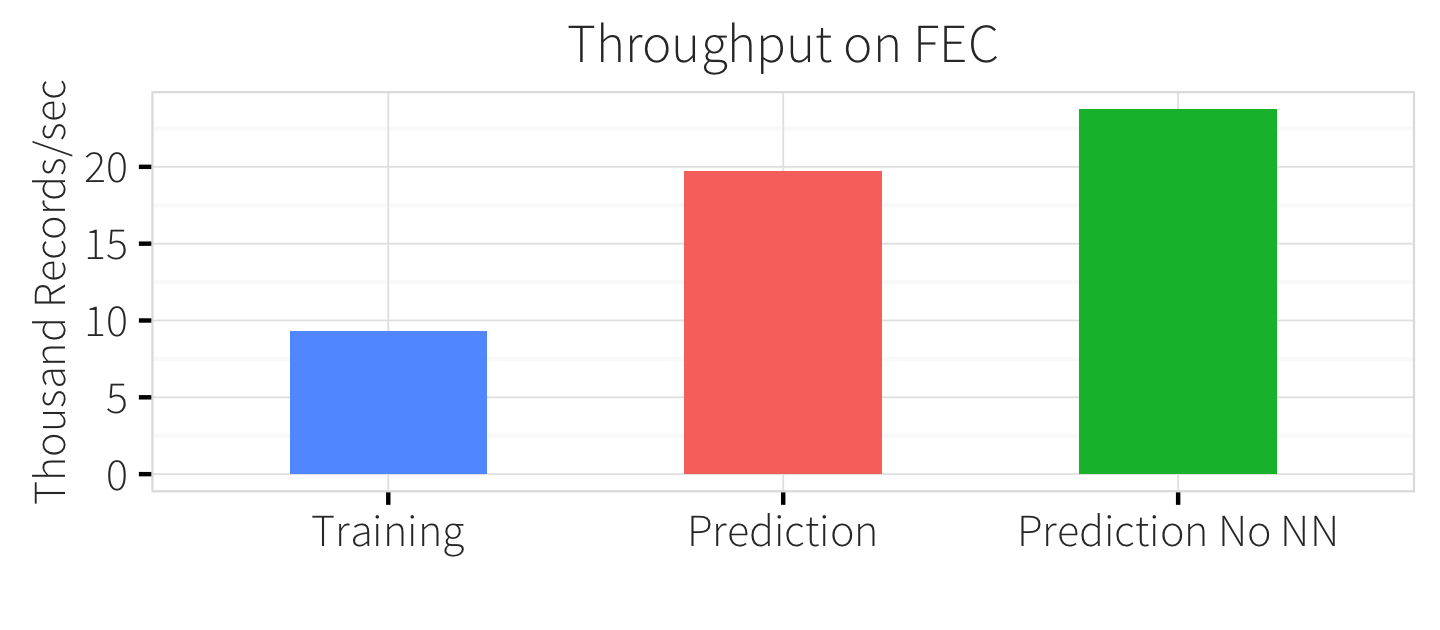}
\caption{Prediction throughput is significantly higher than training throughput (Figure~\ref{exp:runtime}).  Reported for 16-cores.\label{exp:tp}}
\end{figure}

\subsection{End-to-End Run Time}
Next, we evaluate the end-to-end wall clock runtime of \sys. We use the FEC dataset since it is the largest. This evaluation includes all of the optimizations for \sys. The FEC dataset is 1.5 GB (about 6M records). 
Figure \ref{exp:runtime} plots the results.
With a single core, \sys takes 2422 seconds in wall-clock time. Of that time, 2072 seconds is spent in repair selection, 306 seconds is spent in error detection, and 44 seconds in loading the dataset.
We can parallelize the repair selection step. We parallelize the inner-loop of the boosting algorithm. On 16-cores, we are able to reduce the runtime of the repair selection to 212 seconds. This constitutes a 9.7x speedup for that step.

It is important to note that this latency is only incurred during training. During prediction, the learned model can be applied, and this process is much faster than training. 
Figure \ref{exp:tp} plots the throughput of \sys.
The number of records that can be processed per second on 16 cores for prediction is 19746 records/second, but during training it is 9316 records/second. One of the key bottlenecks is evaluating the word2vec model for each prediction, and without this model, the throughput increases to 23746 records/second.

\begin{figure}[t]
\centering
 \includegraphics[width=\columnwidth]{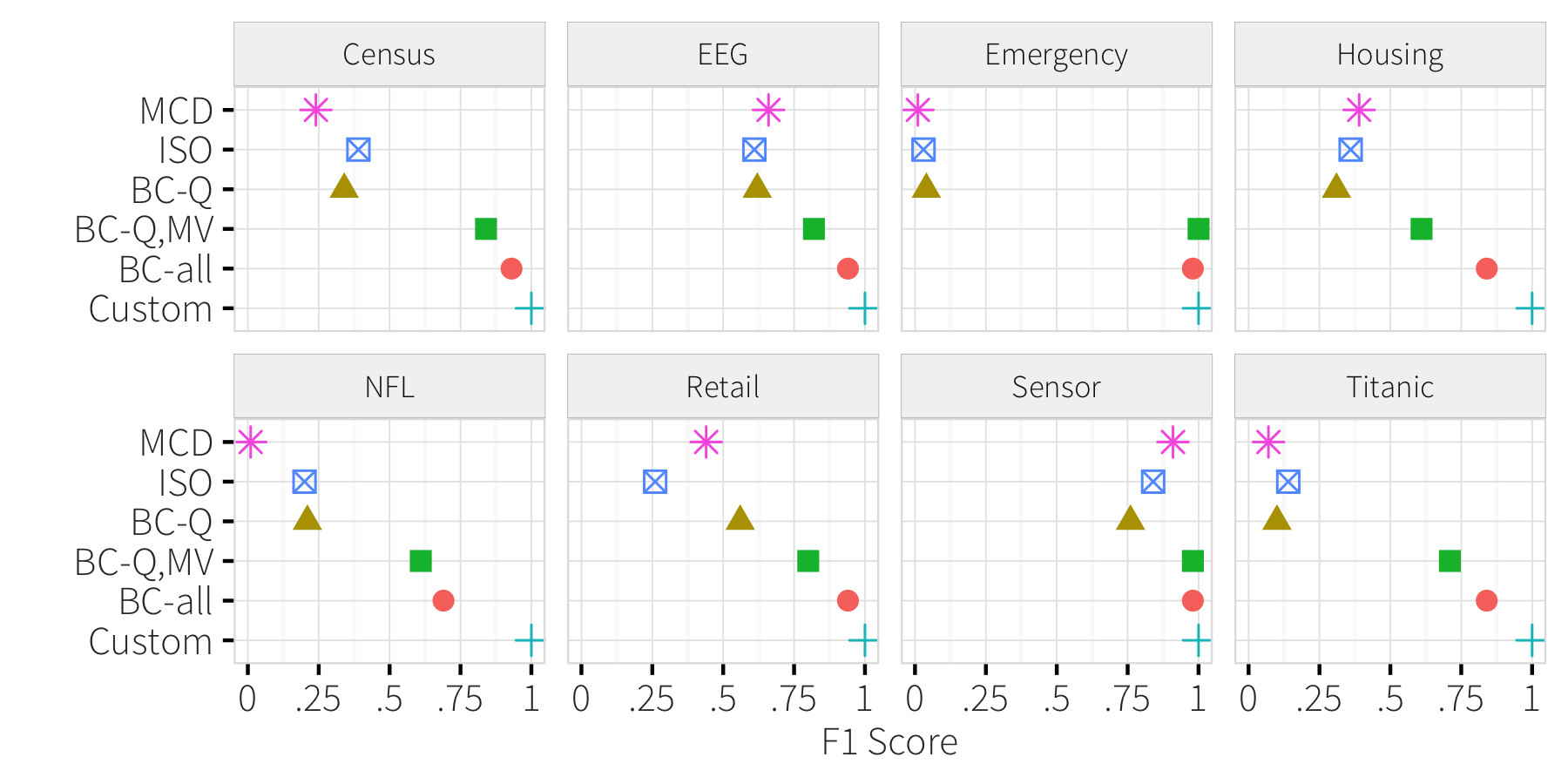}
 \caption{ \sys achieves up to 81\% accuracy and is competitive with hand-written rules, and the word embedding features significantly improve the detector accuracy.
 \label{fig:derror}}
\end{figure}

\begin{figure}[t]
\centering
 \includegraphics[width=\columnwidth]{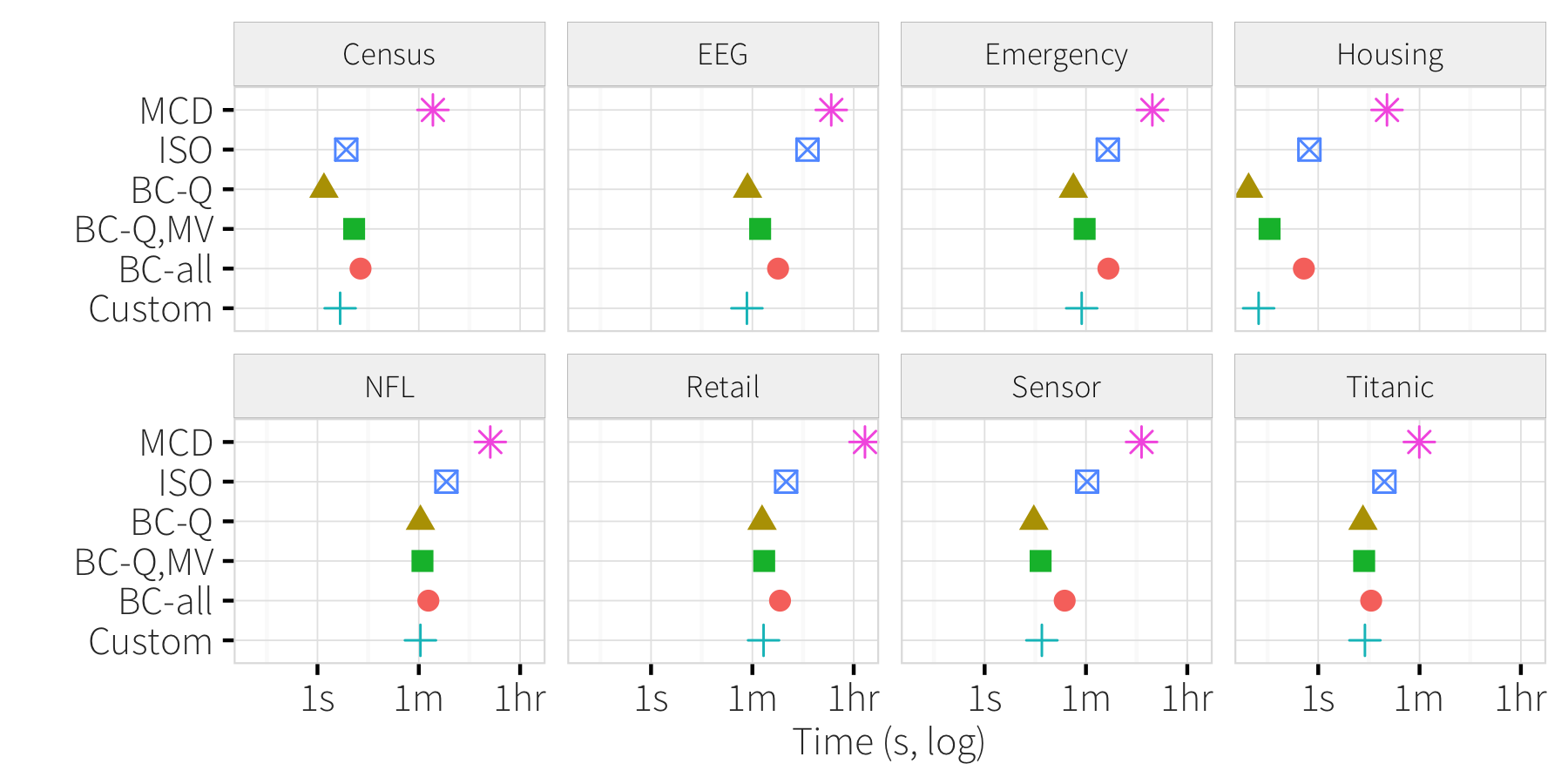}
 \caption{Runtimes for 8 Machine Learning competition datasets as in Figure~\ref{fig:derror}. \sys is slightly slower than hand-written rules, and much faster than MCD.
 \label{fig:druntime}}
\end{figure}

\subsection{Detector Micro-Benchmarks}

We used a set of error detectors based on heuristics, statistical methods, the word2vec neural network, and evaluated their accuracy and runtime as compared to hand crafted rules (Custom).  We evaluated typical outlier detection techniques such as Minimum Covariance Determinants (MCD) and Isolation Forests with naive hot-one encoding for categorical attributes (ISO).  We also evaluated \sys by incrementally the set of featurizers that are used: quantitative only (BC-Q), with  missing value featurizers (BC-Q,MV), with word embeddings (BC-all).  We report the F1 score on the 8 machine learning datasets.

Figure  \ref{fig:derror} shows that the final detector in \sys achieves up to 81\% of the accuracy of hand-written rules on the competition datasets.
Confirming the results of Abedjan et al.~\cite{DBLP:journals/pvldb/AbedjanCDFIOPST16}, we found that a 
purely quantitative approach does not perform well in comparison to the rule-based approach on these datasets (Isolation Forest alone and MCD).
However, results are significantly improved when combined with heuristics that detect missing values. 
The performance gap is even further reduced when the detector additionally uses a Neural Network to learn how attributes correlate with each other, and detect anomolous correlations.
It is important to emphasize that these datasets represent a very specific domain, i.e., structured training datasets for ML.
The datasets are already in a structured schema and the only thing that an analyst has to worry about is handling inconsistent attribute values.
Presumably these datasets were also previously cleaned and extracted before they were publicly released.
Our initial experiment showed that for this class of datasets, reasonably accurate error detection is possible with minimal supervision and tuning.

 Figure  \ref{fig:druntime} shows the total runtime (training and test) of each  approach.
 We first apply the Minimum Covariance Determinant approach (MCD) to the set of records featurized ``naively''--numerical values, hot-one encoded categorical values, and bag-of-words for strings.
 We found that MCD was very expensive since this featurization increases the dimensionality significantly and MCD needs to compute an $d^2$ covariance matrix.
 Next, we apply the Isolation Forest to the same featurization.
 The Isolation forest is up-to $50x$ faster than MCD on the same features.
 This was one of the big motivations for using an isolation forest internally in \sys.
 
 After that, we applied Isolation Forests to progressively more of the featurizers used by \sys.
 First, we applied it just to the numerical attributes--this is the fastest.
 Then, we applied it to numerical attributes, missing values, and parsing errors--this adds $1.5x$ overhead on averaage.
 Finally, we added in the word2vec neural network features (excluding training time for the Neural Network).
 We notice that with this featurization the Isolation Forest is faster than the one with the naive featurization due to the lower dimensionality.
 Of course, rules are faster to evaluate than a learning detector and this gap was on average a factor of 3.
 
 \begin{figure}[t]
\centering
 \includegraphics[width=0.9\columnwidth]{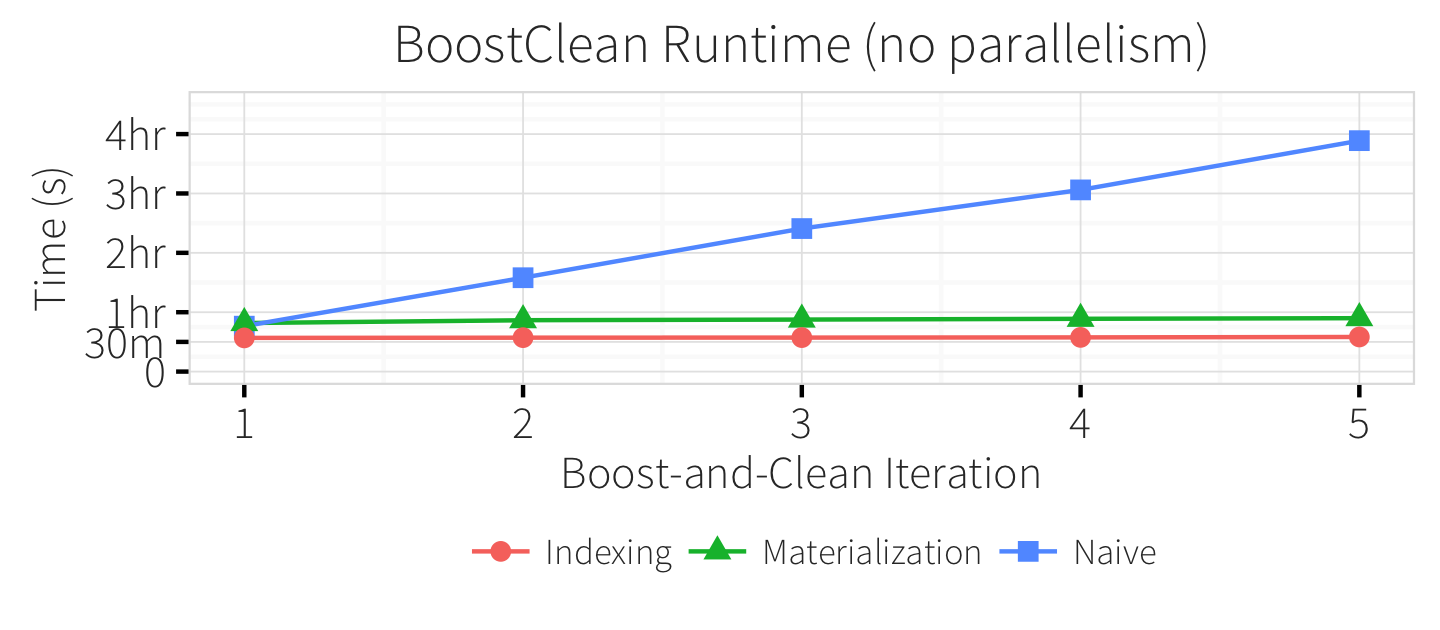}
 \caption{This plot (log scale) shows the impact of optimizations on the selector's runtime. Materialization and Indexing allow the algorithm to scale with the number of selected cleaners. Otherwise, the algorithm repeatedly retrains and recleans the same data.
 \label{fig:opt}}
\end{figure}
 
 \subsection{Repair Micro-Benchmarks }
 
 \stitle{Repair Selector Optimizations:} We proposed two systems optimizations to the boosting algorithm: (1) materialization, and (2) indexing.
 In this set of experiments, we use FEC dataset and apply no parallelism.
 Figure \ref{fig:opt} plots the runtime of the repair selector as a function of the number of cleaners to select (i.e., B).
 Without any optimization, for $B=1$ the repair selector requires 2754 seconds and for $B=5$ requires 14002 seconds.
 The materialization optimization allows us to pay an up-front cost of creating the view during the first iteration of the algorithm, but saves effort on future iterations.
 For $B=1$ with the materialization optimization, the run time is 2943 seconds.
 For $B=5$ the time is drastically cut down to 3241 seconds.
 In each iteration, the indexing algorithm allows us to more efficiently evaluate the accuracy of a cleaner+classifier pair.
 This reduces the run time at $B=5$ to 2072.
 
 \begin{figure}[t]
\centering
 \includegraphics[width=0.9\columnwidth]{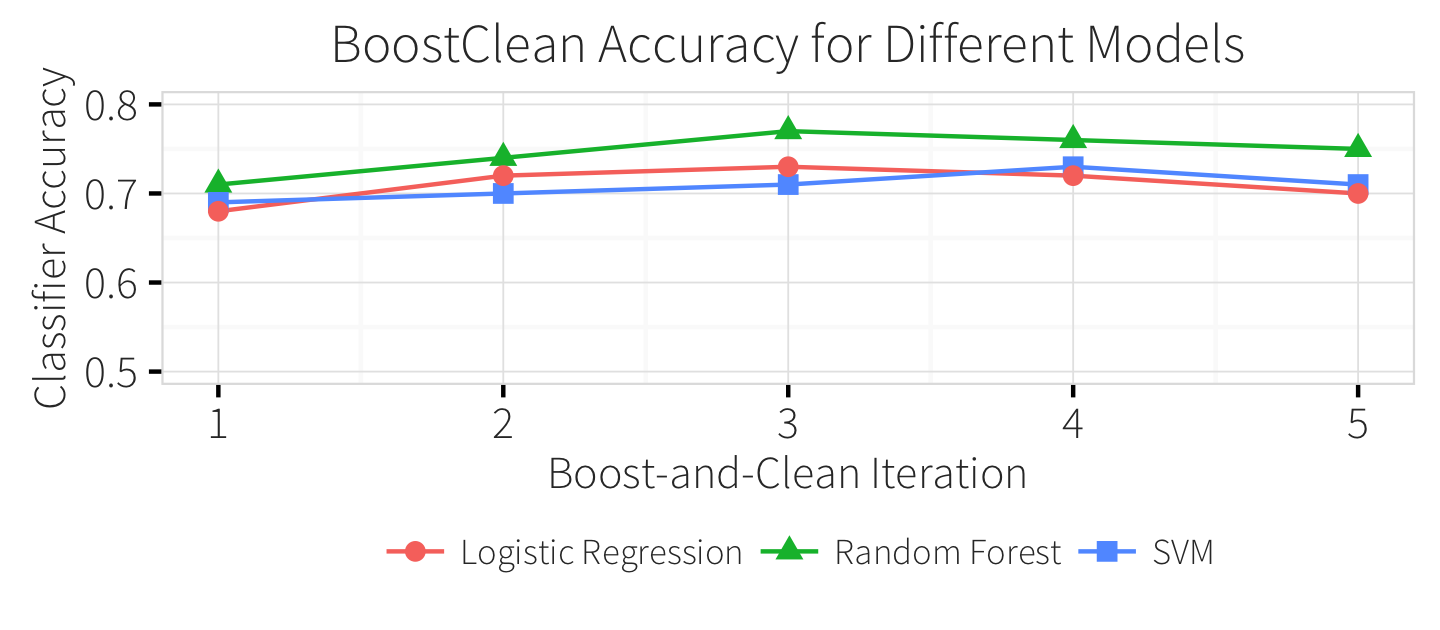}
 \caption{For three different classification models, we plot the learning curves for the repair selector. Selecting too many cleaners can lead to overfitting.
 \label{fig:learning}}
\end{figure}
 
 \stitle{Overfitting: }
 One concern with the repair selector is overfitting. We evaluate to what extent, \sys overfits in Figure \ref{fig:learning}, where we plot the learning curves (accuracy as a function of the number of cleaners $B$).
 We try three different classification models, random forests, SVMs, and logistic regression.
 For all of the models we see similar results, where there is an optimal $B$ to select after which \sys overfits.
This is a major concern on small datasets ($<1000$ records) and with few attributes and (e.g., $<5$). 
For a sufficiently large dataset with proper test and training evaluation, this can be set through cross-validation.

\section{Related Work}

\stitle{Data Cleaning: } Since the beginning of data management, several research and commercial systems have been proposed to improve data cleaning efficiency and accuracy (see~\cite{rahm2000data} for a survey).
Over the past few years, there have been several significant data cleaning advances in scalability~\cite{wang1999sample, khayyat2015bigdansing, altowim2014progressive}.
However, \emph{machine time} is only part of the story and the \emph{human time} of data cleaning is known to be significant.
Several tools have been proposed to reduce human burden including automatically generating explanations~\cite{DBLP:journals/pvldb/0002M13}, building more robust interfaces~\cite{wrangler,trifacta}, and more sophisticated human-in-the-loop processing with crowds~\cite{gokhale2014corleone, park2014crowdfill, DBLP:journals/pvldb/YakoutENOI11, chu2015katara, DBLP:journals/pvldb/HaasKWF015,marcus2015crowdsourced}.
However, due to changes such as the increasing popularity of advanced statistical analytics and the vast amounts of numerical time-series sensor data, many data quality problems that were initially considered for relational analytics may have to be revisited.
We explore how to reduce the burden on data scientists in the context of ML pipelines with clean test labels available.

\stitle{Analysis-Driven Cleaning: } There is a growing body of literature that studies analysis-driven data cleaning, that is, applying data cleaning in a sufficient way to answer a given query.
For example, Altwaijry et al.~\cite{altwaijry2015query} describe a technique for resolving a sufficient subset of entities in a database to answer SPJ queries.
Bergman et al. \cite{DBLP:conf/sigmod/BergmanMNT15} proposed identifying errors in selection query results and generating crowd-scoured queries to determine fixes to the base data.
Similarly, work on the consistent query answering problem explored the minimal effort needed to answer a query given a set of integrity constraints over a dirty relation~\cite{DBLP:series/synthesis/2011Bertossi}.

While the work on relational queries is extensive, analytical queries (aggregates, advanced statistical analytics, learning etc.) is less studied.
Projects like ActiveClean~\cite{DBLP:journals/pvldb/KrishnanWWFG16} have studied algorithms for prioritizing user-defined cleaning using the downstream ML model, ActiveClean does not actually clean the data--it only decides where to apply a predefined operation.
\sys studies an extension where the cleaning operations can be selected from a discrete set given a clean test dataset (can be much smaller) to evaluate the user's analytics.
This approach promises to significantly reduce the effort in designing cleaning software since the time-consuming trial-and-error development process is automated.

\stitle{Machine Learning For Cleaning: } There are a number of other works that use machine learning to improve the efficiency and/or reliability of data cleaning~\cite{DBLP:journals/pvldb/YakoutENOI11,yakout2013don,gokhale2014corleone}.
For example, Yakout et al. train a model that evaluates the likelihood of a proposed replacement value \cite{yakout2013don}.
Another application of machine learning is value imputation, where a missing value is predicted based on those records without missing values.
Machine learning is also increasingly applied to make automated repairs more reliable with human validation \cite{DBLP:journals/pvldb/YakoutENOI11}.
Human input is often expensive and impractical to apply to entire large datasets.
Machine learning can extrapolate rules from a small set of examples cleaned by a human (or humans) to uncleaned data \cite{gokhale2014corleone, DBLP:journals/pvldb/YakoutENOI11}.
This approach can be coupled with active learning \cite{DBLP:journals/pvldb/MozafariSFJM14} to learn an accurate model with the fewest possible number of examples.
While, in spirit, \sys is similar to these approaches, it addresses a very different problem of data cleaning optimization for user-defined ML-based analytics.

\stitle{Alternative Learning Models: } Furthermore, there are alternative ensembling approaches that could be considered like Multi-Arm Bandits~\cite{bubeck2013multiple}. 
In our particular problem statement, we assume a fixed test set.
This means that the problem is deterministic unlike the bandit setting.
Furthermore, we are interested in selecting a subset that jointly maximizes prediction accuracy and not a top-k.
We hope to explore this avenue in the future and this might be promising for ``weak'' accuracy metrics.

 \section{Conclusion and Future Work}
We have shown that automated data cleaning for predictive models can be cast in a statistical boosting framework.  We have prototyped this idea in \sys, a new data cleaning system that detects errors in ML data and uses knowledge of the labels to adaptively select from a set of repair actions to maximize prediction accuracy.
We evaluated results on 8 ML datasets on Kaggle and the UCI repository with real data errors and compare to statistical anomaly detection techniques, constraint-based techniques, and the best single cleaner performance. In all 8 datasets, \sys increased the  test accuracy over alternatives. In addition, we evaluated \sys on production datasets from a data science company and showed that, despite high class imbalances in both datasets, \sys can automatically detect data errors and improve the AUC of the downstream model by $8-9\%$.  We also demonstrate how our optimizations can achieve an end-to-end speed up of over $22\times$

We are excited about these promising results and have identified a number of future research directions to improve the practicality of the system.  The first is to relax the current requirements of having a test set with clean labels.   Although it may be difficult to acquire sufficient test labels, data science application often have access to an indirect model accuracy measure.  For instance, user retention may be strongly correlated to model accuracy and much easier to obtain.  This will likely require a more complex ensembling technique than boosting.  A second direction is to support parameterized cleaning operations, such as regular expression extractors, for which the number of possible parameter values is unbounded.  We believe that recursive discretization procedures are a promising approach.  A third direction are further performance optimizations so that \sys can scale to large and heterogeneous settings such as data lakes.  Finally, we are actively seeking to continue industrial collaborations and real-world evaluations of our system.

\bibliographystyle{abbrv}
\bibliography{main} 
\appendix

\vspace{0.5em}\noindent\textbf{USCensus: } This dataset contains US Census records for adults and the goal is to predict  whether the adult earns more than $50,000$ dollars. It contains 32,561 records with 15 numerical and categorical attributes. This dataset contained missing values and coding inconsistencies.
Examples of data error include:
\begin{lstlisting}
#missing values
40,Private,121772,Assoc-voc,11,
Married-civ-spouse,Craft-repair,Husband, 
Asian-Pac-Islander,Male,0,0,40,(*\orange{\bf{?}}*),>50K

#coding inconsistency
57,Local-gov,110417,HS-grad,9,
Married-civ-spouse,Craft-repair,Husband,
White,Male,(*\orange{\bf{99999}}*),0,40,United-States,>50K
\end{lstlisting}

\vspace{0.5em}\noindent\textbf{NFL: } This dataset contains play-by-play logs from US Football games. The dataset contains 46,129 records with 65 numerical, categorical, and string-valued attributes. Given the record, the classification objective is to determine whether the next play the team runs is a run or a pass play.
The dataset contains a significant number of missing values and ``sentinel'' records that mark the end of a log sequence. The sentinel records do not signify a play but rather signify a timeout, end of quarter, or end of the game.
\begin{lstlisting}
#missing values
"36",2015-09-10,"2015091000",1,1,(*\orange{\bf{NA}}*),"15:00",
15,3600,0,"NE",35,35,0,0,0,(*\orange{\bf{NA}}*),"PIT","NE"(*\blue{\bf{....}}*)

#sentinel record
"189710",2016-01-03,"2016010310",10,2,NA,"00:00",
0,1800,8,"GB",17,17,0,-1,0,0,"",NA,"END(*~*)QUARTER2"
,1,0,0,0,NA, NA,NA,0,"Quarter(*~*)End"(*\blue{\bf{....}}*)
\end{lstlisting}

\vspace{0.5em}\noindent\textbf{EEG: } This is a dataset of EEG recordings. 
The training data is organized into ten minute EEG clips labeled "Preictal" for pre-seizure data segments, or "Interictal" for non-seizure data segments. 
There are 2406 records each of which is a variable-length time-series of 16 attributes. We featurize this dataset into records of 32 attributes--the mean and variance over the length of the time-series. 
This dataset primarily contains numerical outliers, the clips have spurious readings.
\begin{lstlisting}
#Time t=46 Normal
[-41.53080368041992, -9.605541229248047, 
-55.74542999267578, 17.77084732055664,
-1.6866581439971924, 38.86453628540039, 
17.108707427978516, 26.545927047729492, 
-12.696817398071289, -12.703478813171387, 
56.78707504272461, 3.2556533813476562, 
22.688213348388672, -25.728403091430664, 
-10.142332077026367, -11.585281372070312]

#Time t=47 Abnormal
(*\orange{[0, 8, -10, 9, 18, 6, -8, -41, -26, -72, -19, 70, 129, 53, 31, -11]}*)
\end{lstlisting}

\vspace{0.5em}\noindent\textbf{Sensor: } The Intel sensor dataset contains 928,991 temperature, humidity, and light sensor readings a sensor deployment. The classification task is to predict whether the readings came from a particular sensor (sensor 49). This dataset primarily has numerical outliers.
\begin{lstlisting}
#Normal Record
49  -0.999750  12.862100  10.368300  10.438300  
11.669900 (*\orange{\bf{13.493100}}*)  13.342300  8.041690  
8.739010  26.225700  59.052800

#Spurious Record
49  1.175188  12.279100  8.849360  9.005830  
10.111700  (*\orange{\bf{378.750000}}*)  19.319400  15.916200  
37.631400  27.150100  53.403700
\end{lstlisting}

\vspace{0.5em}\noindent\textbf{Titanic: } This dataset contains 891 records from the Titanic manifest with 12 attributes. The classification objective is to determine whether the passenger survived or not. There are missing values and string formatting errors.

\begin{lstlisting}
#missing values
891,0,3,"Dooley, Mr. Patrick",male,
32,0,0,370376,7.75,(*\orange{\bf{--}}*),Q
\end{lstlisting}

\vspace{0.5em}\noindent\textbf{Housing: } The housing dataset contains 1460 records and 81 attributes of house price listings. The classification objective is to determine whether the listed house will be sold above 750000. 
This dataset contains missing values as well as numerical outliers.
\begin{lstlisting}
#missing values
(*\blue{\bf{....}}*)204,228,0,0,0,(*\orange{\bf{NA,NA}}*),Shed,350,11,2009,WD,
Normal,200000
\end{lstlisting}

\vspace{0.5em}\noindent\textbf{Retail: } The online retail dataset contains 541,909 records of online retail purchases with 8 attributes. The classification objective is to predict whether the purchase occurred in the United Kingdom.
This dataset contains numerical errors where some purchased quantities are reported as negative.

\begin{lstlisting}
#outliers
C536391,21980,PACK OF 12 RED RETROSPOT TISSUES
,(*\orange{\bf{-24}}*),12/1/10 10:24,0.29,17548,United Kingdom
\end{lstlisting}

\vspace{0.5em}\noindent\textbf{Federal Election Commission Contributions: } The FEC provides a dataset of election contributions of 6,410,678 records with 18 numerical, categorical and string valued attributes. This dataset has a number of errors. There are missing values, formatting issues (where records have the wrong number of fields causing misaligment in parsing), and numerical outliers (negative contributions).

\begin{lstlisting}
#missing values
C00458844,"P60006723","Rubio, Marco","RUCINSKI,
ROBERT","APO","AE","090960009","US ARMY",
"PHYSICIAN",100,08-MAR-16,(*\orange{\bf{``''}}*),(*\orange{\bf{``''}}*),(*\orange{\bf{``''}}*),"SA17A",
"1082559","SA17.1074981","P2016"

#rejected contributions double recorded
C00458844,"P60006723","Rubio, Marco","SWAID, 
SWAID N. DR.","BIRMINGHAM","AL","352660827",
"NEWOLOGICAL SURGERY ASSOCIATES","PHYSICIAN",
(*\orange{\bf{-400}}*),28-DEC-15, "REDESIGNATION TO GENERAL","X",
"REDESIGNATION TO GENERAL","SA17A",
"1047126","SA17.892835B","P2016"
\end{lstlisting}

\vspace{0.5em}\noindent\textbf{Restaurant Dataset: } The restaurant dataset has 758 distinct records and 4 attributes. This dataset has typically been used as a benchmark for entity resolution since records are duplicated with minor inconsistencies.
We designed a multi-class classification task to see if we could predict the city from record.
One of the major inconsistencies was additional attributes appended to the restaurant category.

\begin{lstlisting}
campanile,624 s. la brea ave.,los angeles,
american

grill  the,9560 dayton way,beverly hills,
american (*\orange{\bf{(traditional)}}*)
\end{lstlisting}

\vspace{0.5em}\noindent\textbf{Housing: } The housing dataset contains 1460 records and 81 attributes of house price listings. The classification objective is to determine whether the listed house will be sold above 750000. 
This dataset contains missing values as well as numerical outliers.

\begin{lstlisting}
#missing values
(*\blue{\bf{....}}*)204,228,0,0,0,(*\orange{\bf{NA,NA}}*),Shed,350,11,2009,WD,
Normal,200000
\end{lstlisting}

\end{document}